\documentclass[twocolumn]{aastex631}
\usepackage{subfigure}
\usepackage{amsmath}
\usepackage{autobreak}
\usepackage{hyperref}
\hypersetup{
    colorlinks=true,
    citecolor = blue,
    linkcolor=blue,
    urlcolor=blue,
    pdfpagemode=FullScreen,
    }

\graphicspath{{./}{figures/}}
\begin{document}
\title{Investigating 16 Open Clusters in the Kepler/K2 - Gaia DR3 field. I. Membership, Binary, and Rotation}
\author[0000-0003-2908-1492]{Liu Long}
\affiliation{Institute for Frontiers in Astronomy and Astrophysics, Beijing Normal University,  Beijing 102206, China}
\affiliation{Department of Astronomy, Beijing Normal University, Beijing 100875, People’s Republic of China}

\author[0000-0002-7642-7583]{Shaolan Bi}
\affiliation{Institute for Frontiers in Astronomy and Astrophysics, Beijing Normal University, Beijing 102206, China}
\affiliation{Department of Astronomy, Beijing Normal University, Beijing 100875, People’s Republic of China}
\email{bisl@bnu.edu.cn}

\author[0000-0002-2510-6931]{Jinghua Zhang}
\affiliation{South-Western Institute for Astronomy Research, Yunnan University, Chenggong District, Kunming 650500, People's Republic of China}
\email{zhjh@nao.cas.cn}

\author[0000-0002-3672-2166]{Xianfei Zhang}
\affiliation{Institute for Frontiers in Astronomy and Astrophysics, Beijing Normal University,  Beijing 102206, China}

\affiliation{Department of Astronomy, Beijing Normal University, Beijing 100875, People’s Republic of China}

\author[0000-0002-2394-9521]{Liyun Zhang}
\affiliation{College of Physics, Guizhou University, Guiyang 550025, People’s Republic of China}
\affiliation{
State Key Laboratory of Public Big Data, Guizhou University, Guiyang 550025, China}
\author[0000-0002-2614-5959]{Zhishuai Ge}
\affiliation{Beijing Planetarium, Beijing Academy of Science and Technology, Beijing, 100044, China}
\author[0000-0001-6396-2563]{Tanda Li}
\affiliation{Institute for Frontiers in Astronomy and Astrophysics, Beijing Normal University,  Beijing 102206, China}
\affiliation{Department of Astronomy, Beijing Normal University, Beijing 100875, People’s Republic of China}

\author[0000-0003-3957-9067]{Xunzhou Chen}
\affiliation{Research Center for Intelligent Computing Platforms, Zhejiang Laboratory, Hangzhou 311100, China}
\author[0000-0003-3020-4437]{YaGuang Li}
\affiliation{Sydney Institute for Astronomy (SIfA), School of Physics, University of Sydney, NSW 2006, Australia}
\affiliation{Stellar Astrophysics Centre, Department of Physics and Astronomy, Aarhus University, Ny Munkegade 120, DK-8000 Aarhus C, Denmark}

\author{LiFei Ye}
\affiliation{Institute for Frontiers in Astronomy and Astrophysics, Beijing Normal University,  Beijing 102206, China}
\affiliation{Department of Astronomy, Beijing Normal University, Beijing 100875, People’s Republic of China}

\author[0000-0003-0795-4854]{TianCheng Sun}
\affiliation{Institute for Frontiers in Astronomy and Astrophysics, Beijing Normal University,  Beijing 102206, China}
\affiliation{Department of Astronomy, Beijing Normal University, Beijing 100875, People’s Republic of China}

\author{JianZhao Zhou}
\affiliation{Institute for Frontiers in Astronomy and Astrophysics, Beijing Normal University,  Beijing 102206, China}
\affiliation{Department of Astronomy, Beijing Normal University, Beijing 100875, People’s Republic of China}

\begin{abstract}
Using data from the Gaia Data Release 3 (Gaia DR3) and Kepler/K2, we present a catalog of 16 open clusters with ages ranging from 4 to 4000 Myr, which provides detailed information on membership, binary systems, and rotation. We assess the memberships in 5D phase space, and estimate the basic parameters of each cluster. Among the 20,160 members, there are 4,381 stars identified as binary candidates and 49 stars as blue straggler stars. The fraction of binaries vary in each cluster, and the range between 9\% to 44\%. We obtain the rotation periods of 5,467 members, of which 4,304 are determined in this work. To establish a benchmark for the rotation-age-color relation, we construct color-period diagrams. We find that the rotational features of binaries are similar to that of single stars, while features for binaries are more scattered in the rotation period. Moreover, the morphology of the color-period relationship is already established for Upper Scorpius at the age of 19 Myr, and some stars of varying spectral types (i.e. FG-, K-, and M-type) show different spin-down rates after the age of $\sim$ 110 Myr. By incorporating the effects of stalled spin-down into our analysis, we develop an empirical rotation-age-color relation, which is valid with ages between 700 - 4000 Myr and colors corresponding to a range of 0.5 $<$ $(G_{BP}-G_{RP})_0$ $<$ 2.5 mag.

\end{abstract}
\keywords{\href{http://astrothesaurus.org/uat/1567}{Star clusters (1567)}; \href{http://astrothesaurus.org/uat/1160}{Open star clusters (1160)}; \href{http://astrothesaurus.org/uat/1629}{Stellar rotation (1629)}; \href{http://astrothesaurus.org/uat/1581}{Stellar ages (1581)}; \href{http://astrothesaurus.org/uat/154}{Binary stars (154)}}
\section{Introduction}\label{sec:intro}
\indent Open clusters (OCs) are benchmarks for comprehending star and stellar evolution \citep{Lada2003ARA&A..41...57L,Portegies2010ARA&A..48..431P,2019ARA&A..57..227K}, as the members in individual OC form contemporaneously and share the same age, kinematics, chemical abundance, and distance. Additionally, OCs provide the ideal environment for studying binary or multiple systems. In the dense environment of the OCs, binaries are closely related to the dynamic evolution of OCs. \citep{Heggie1975MNRAS.173..729H,Kaczmarek2011A&A...528A.144K,Grijs2015arXiv151000099D,Deacon2020MNRAS.496.5176D}. Particularly, the evolution of binary stars in OCs can produce special stellar objects like blue straggler star (BSS)\citep{Perets2015ASSL..413..251P,St2017A&A...597A..87S}. The merger or mass transfer between two components in a binary system can potentially lead to the formation of BSS \citep{McCrea1964MNRAS.128..147M,Andronov2006ApJ...646.1160A,Geller2011Natur.478..356G}. Therefore, studying BSS in OCs provides valuable insights into the dynamical processes of the clusters, and the contribution of binaries to cluster evolution \citep[e.g.][]{Bailyn1995ARA&A..33..133B, Ferraro2012Natur.492..393F, Dresbach2022ApJ...928...47D,Jiang2022ApJ...940...97J}.\\
\indent Stellar rotation plays a crucial role in the stellar structure and evolution. The observations of stellar rotation in open clusters from young to old could provide us information about stellar internal behaviors at different evolutionary stages. \citet{Skumanich1972ApJ...171..565S} was the first to discover stellar rotation rate decrease with age in Sun-like stars, and this relation can be described by a simple power law. \citet{Barnes2003} subsequently extended the rotation-age relation to late-type stars in other colors and discovered a tight relation among rotation, age, and color (or mass). This relation can be used to infer stellar ages by measuring the rotation period and color(or mass) of stars, which is known as gyrochronology.
The extensive works are committed to clarifying the accuracy of this method and to exploring the underlying physical processes of gyrochronology \citep[e.g.][]{Barnes2010, Mamajek2008ApJ...687.1264M, Matt2012ApJ...754L..26M,Matt2015ApJ...799L..23M,Gallet2015A&A...577A..98G,van2013ApJ...776...67V,van2016Natur.529..181V,Angus2019AJ....158..173A,Spada2020A&A...636A..76S, Kounkel2022AJ....164..137K,Otani2022ApJ...930...36O}. However, the current models fail to fully reproduce the observed rotation sequences of OCs. This discrepancy is particularly prominent among K dwarfs, which stars exhibit a temporary stall of rotation in clusters older than approximately 1 Gyr \citep{Agueros2018ApJ...862...33A,Curtis2019ApJ...879...49C, Curtis2020ApJ...904..140C}. Moreover, limited observational data for clusters older than 1 Gyr restricts the validation and calibration of gyrochronology relations in the older age. Additionally, investigating the behavior of binary systems is critical for a better understanding of the rotational evolution because binary components have the same ages \citep{Chanam2012ApJ...746..102C,Messina2019A&A...627A..97M,Simonian2019ApJ...871..174S,Otani2022ApJ...930...36O}. Nonetheless, it remains uncertain whether the rotational evolution of binary systems follows the same rules as that of single stars.\\
\indent Thanks to the high-precision and long-term observations provided by Kepler\citep{Borucki2010Sci...327..977B} and K2\citep{Howell2014PASP..126..398H}, the rotation periods of tens of thousands of stars have been measured \citep{McQuillan2013MNRAS.432.1203M,Santos2021ApJS..255...17S,Reinhold2020}. One of the advantages of Kepler and K2 missions is the inclusion of OC in its target list, particularly K2, which has observed over a dozen clusters (such as Pleiades, Praesepe, M67). The independent studies of these clusters have been carried out by previously published (e.g. Taurus \citep{Rebull2020}, $\rho$ Ophiuchus and Upper Scorpius \citep{Rebull2017}, M35 and Pleiades \citep{Meibom2009}, Hyades and Praesepe \citep{Delorme2011MNRAS.413.2218D,Douglas2016,Douglas2019} etc.). 
While prior studies have extensively examined the rotation of these clusters, they have primarily focused on individual cluster or several clusters of similar ages. Our main purpose is to construct a sample of star clusters ranging from young to old, enabling the investigation of rotation distribution characteristics across various ages and spectral types, including single stars and binary systems.
With the advent of precise astrometric data from the Gaia survey, photometric data with unprecedented high-precision kinematics and parallax are provided \citep{Gaia2022arXiv220800211G}. The accuracy of Gaia DR3 and the increase of observation sources can identify more members and clean the polluted stars in the OCs. By combining the Gaia DR3 and Kepler/K2 data, we aim to investigate the membership, binary, stellar rotation, and basic parameters of 16 clusters with the intent of providing an updated view of stellar rotation-age-color relation. \\
\indent In Section \ref{sec:CM}, we display the sample of 16 OCs in Kepler/K2, and describe the method for identifying cluster members using Gaia DR3 astrometry. We estimate the fundamental properties of the identified clusters in Section \ref{sec:Properties}. In Section \ref{sec:binary}, we identify binary candidates and calculate binary fractions in 16 OCs. In Section \ref{sec:Rotation}, we determine the rotation periods of membership catalogs, and present period-color diagrams. The conclusions are followed in Section \ref{sec:conclusion}.\\
\section{Cluster membership}\label{sec:CM}
\subsection{Cluster samples}\label{sec:sample}
The original $Kepler$ mission contained four OCs (NGC 6866, NGC 6811, NGC 6819 and NGC 6971). $K2$, the extended $Kepler$ mission, provide a wealth of photometric data for nearby pre-main sequence (PMS) stars in young associations (Tuarus, NGC 6530, $\rho$ Ophiuchus and Upper Scorpius) and several OCs (M35, Pleiades, Hyades, Praesepe, NGC 1647, NGC 1750, NGC 1758, NGC 1817, NGC 2158, NGC 6774, M67) \citep{Cody2018RNAAS...2..199C}. Of the clusters observed by $Kepler$ and $K2$, we select two forming associations (Taurus, Upper Scorpius) and 14 OCs ($\rho$ Ophiuchus, M35, Pleiades, Hyades, Praesepe, NGC 1647, NGC 1750, NGC 1758, NGC 1817, NGC 6774, M67, NGC 6866, NGC 6811, NGC 6819) as cluster samples. We exclude clusters that have too few stars with rotation period detections in Kepler and K2 fields, such as NGC 2158 and NGC 6791. The compilation of clusters is provided in Table \ref{tab:sample}. 
\begin{table}[h!t]
\scriptsize
   \centering
   \caption{Associations and OCs observed during the $Kepler$ and $K2$ Missions}
   \label{tab:sample}
   \begin{tabular}{l l c c}
   \hline
   \hline
	Cluster& Campaign & Type\\ 
	\hline
	NGC 6819                &    Kepler prime     &  oc         \\
NGC 6811                &    Kepler prime     &  oc         \\
NGC 6866                &    Kepler prime     &  oc         \\
M35 (NGC 2168)          &    K2 C0            &  oc         \\
Upper Scopius           &    K2 C2, C15       &  association\\
$\rho$ Ophiuchus        &    K2 C2            &  oc         \\
Pleiades (Melotte 22)   &    K2 C4            &  oc         \\
Hyades (Melotte 25)     &    K2 C4, C13       &  oc         \\
M67 (NGC 2682)          &    K2 C5, C16, C18  &  oc         \\
Praesepe (NGC 2632)     &    K2 C5, C16, C18  &  oc         \\
NGC 6774 (Ruprecht 147) &    K2 C7            &  oc         \\
NGC 1647                &    K2 C13           &  oc         \\
NGC 1750                &    K2 C13           &  oc         \\
NGC 1758                &    K2 C13           &  oc         \\
NGC 1817                &    K2 C13           &  oc         \\
Taurus                  &    K2 C13           & association \\
     \hline
    \end{tabular}
\end{table}

\subsection{Memberships}\label{sec:Membership}
We identify the cluster membership by following steps: (1) selecting the sample sources and (2) calculating the membership probability. We take M67 as a working example to describe each step in detail.
\subsubsection{Selecting the sample sources}
\begin{figure}[ht!]
\plotone{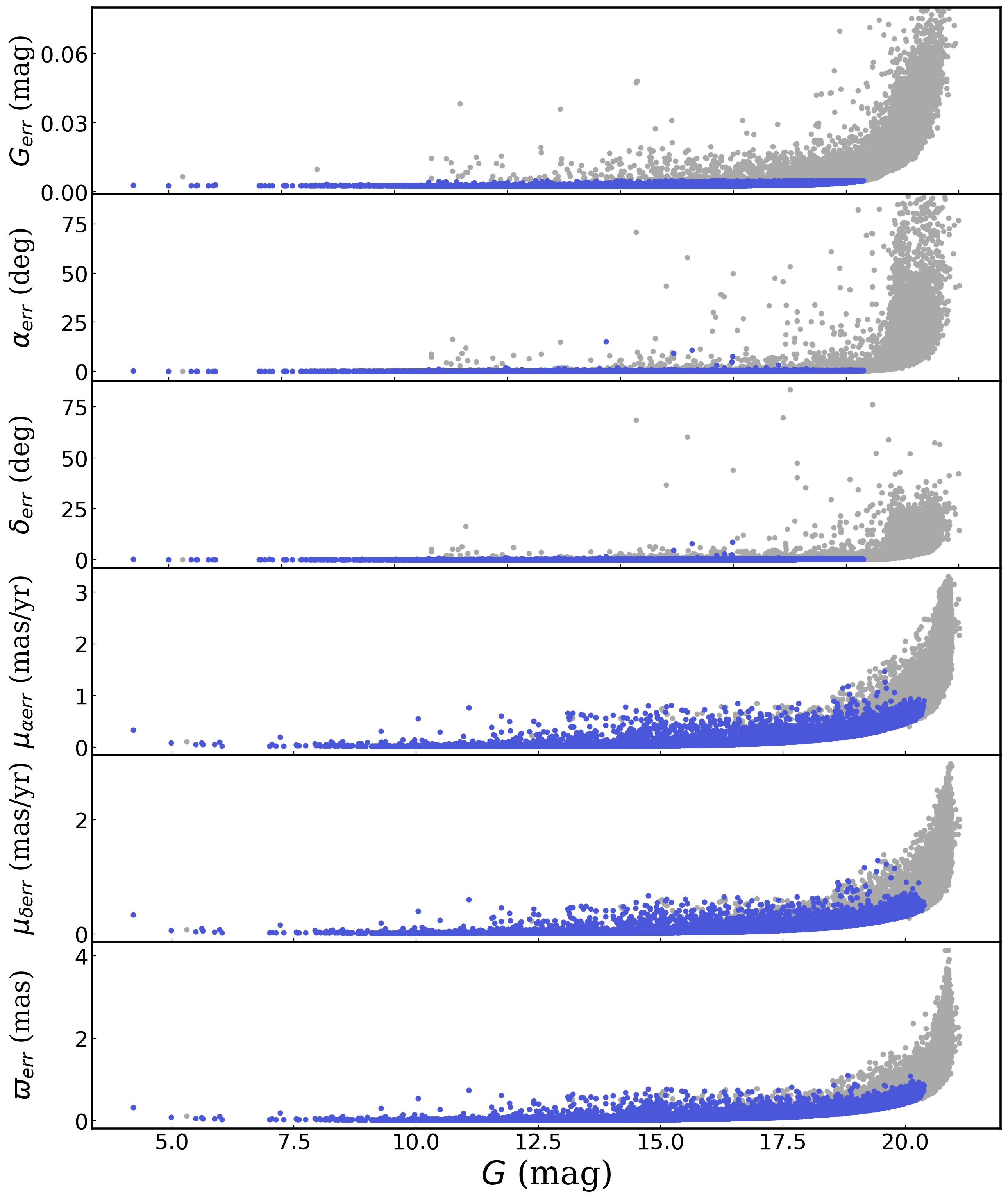}
\caption{The correlation of errors (photometry, parallax, and proper motions) with the magnitude of $G$. The gray points are all $Gaia$ DR3 for M67 within a radius of 10 degrees from the cluster centre. The sources in blue are those with $G_{err}$ $\leq$ 0.005 mag.} \label{fig:sample_err}
\end{figure}

\begin{figure}[ht!]
\plotone{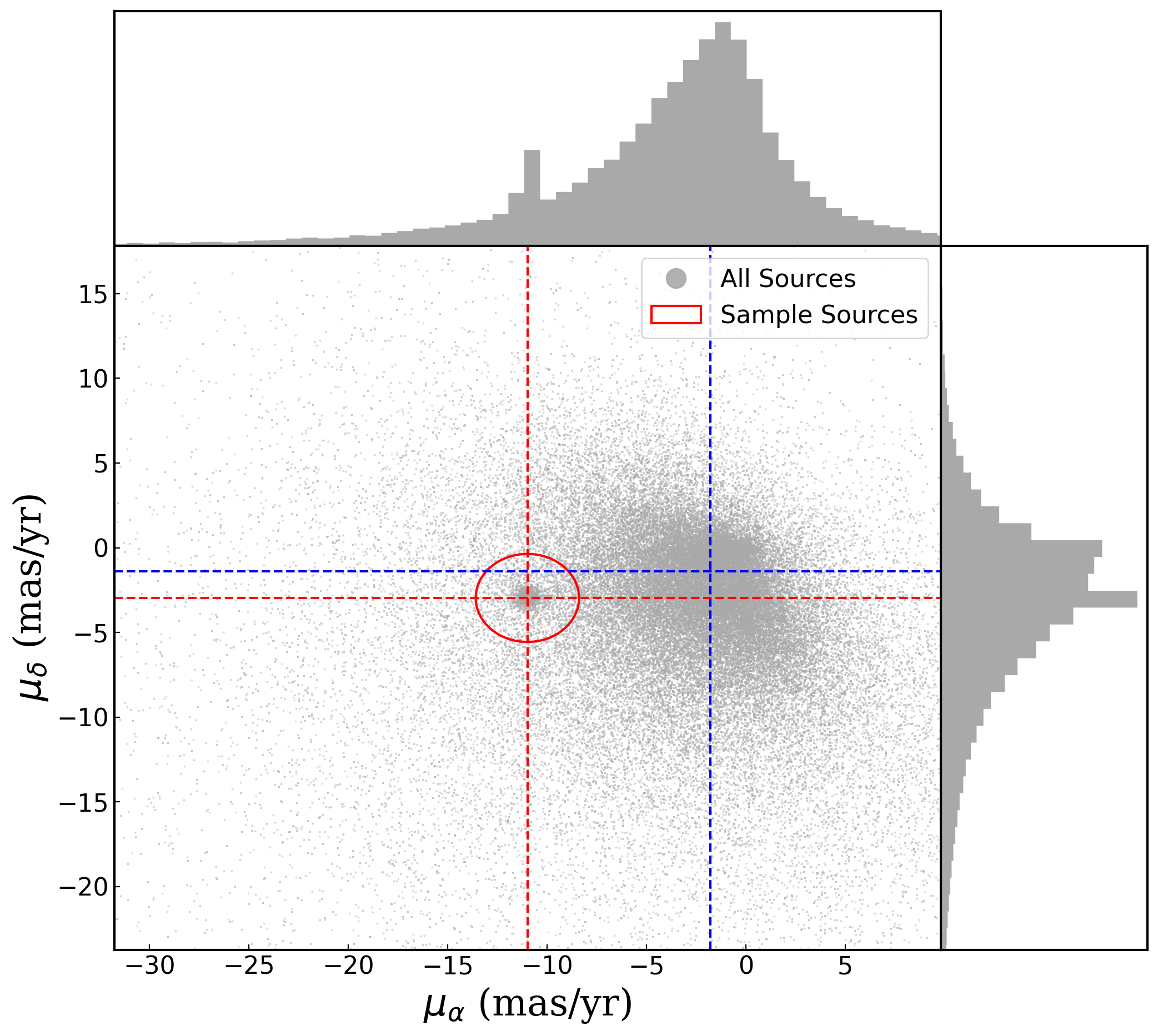}
\caption{The vector point diagram (VPD) of the M67. The gray points are all sources of M67. The sources inside the red circle are $Sample$ $sources$ of M67. \label{fig:M67VPD}}
\end{figure}

According to the method in \cite{2021MNRAS.502.2582A}, we first download sources from Gaia DR3 catalog \footnote{\url{http://tapvizier.u-strasbg.fr/adql/?gaiadr3}} centered on the referred cluster centre coordinates with a radius that is greater than the tidal radius by 1.5 times to adequately include members. The centre coordinates and tidal radius for each OC are taken from \citet{2013A&A...558A..53K}. Then, we choose the sources that satisfy the following criteria:
\begin{enumerate}
\item For each source, the following five astrometric parameters must be included, which are the positions($\alpha$, $\delta$), proper motions($\mu_{\alpha}$, $\mu_{\delta}$), and parallaxes($\varpi$), as well as three photometric pass-bands: $G$, $G_{BP}$, and $G_{RP}$ in the Gaia DR3 catalogue.
\item The accuracy of the astrometric parameters depends on the signal-to-noise ratio, which is related to the magnitude of $G$. Fig.\ref{fig:sample_err} shows the errors in positions, proper motions, and parallax against the magnitude of $G$. We set the errors in the magnitude of $G$ less than 0.005 mag i.e., $G_{err}$ $\leq$ 0.005, which not only removes sources with high uncertainty but also still retaining a part of sources down to $G$ $\backsim$ 21 mag.
\end{enumerate}
 Based on the above criteria, 68,883 sources are remained for M67, which we referred as '$All$  $sources$'. \\
\indent To filter out contamination by a large number of field stars, we need to clip the appropriate window of proper motions that contains all possible cluster members. A rough sample of cluster members is firstly obtained with a clustering algorithm, namely, Hierarchical Density-Based Spatial Clustering of Applications with Noise \citep[HDBSCAN;][]{Campello,McInnes2017JOSS....2..205M}, by identifying over-densities in 3D phase space ($\mu_{\alpha}$, $\mu_{\delta}$ and $\varpi$) using a scaling within (-1, 1) and setting {\tt min$\_$cluster$\_$size} = 4. We use this sample to estimate the mean proper motions and the radius of proper motions (marked as $R_{pm}$) of the cluster. Then, we set windows of proper motions to be a circle centered on the mean proper motions estimated with HDBSCAN and a radius twice that of $R_{pm}$. The sources that are clipped by the proper motion windows are referred as '$Sample$ $sources$'. We obtain 3176 '$Sample$ $sources$' for M67. Fig.\ref{fig:M67VPD} shows the vector point diagram (VPD) of the '$Sample$ $sources$' and '$All$ $sources$' of the cluster.

\subsubsection{Identifying the members}

\begin{figure}[ht!]
\plotone{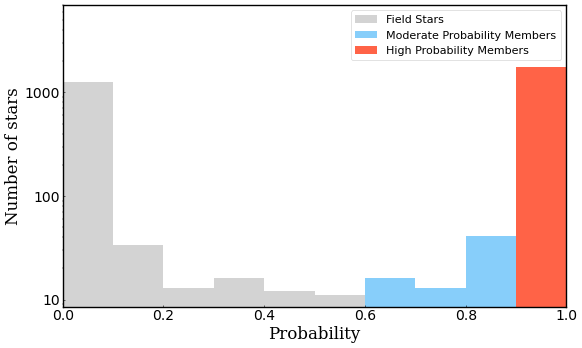}
\caption{Distribution of the membership probabilities for the $Sample$ $sources$ in the M67 field.\label{fig:M67_prob}}
\end{figure}

\begin{figure}[ht!]
\plotone{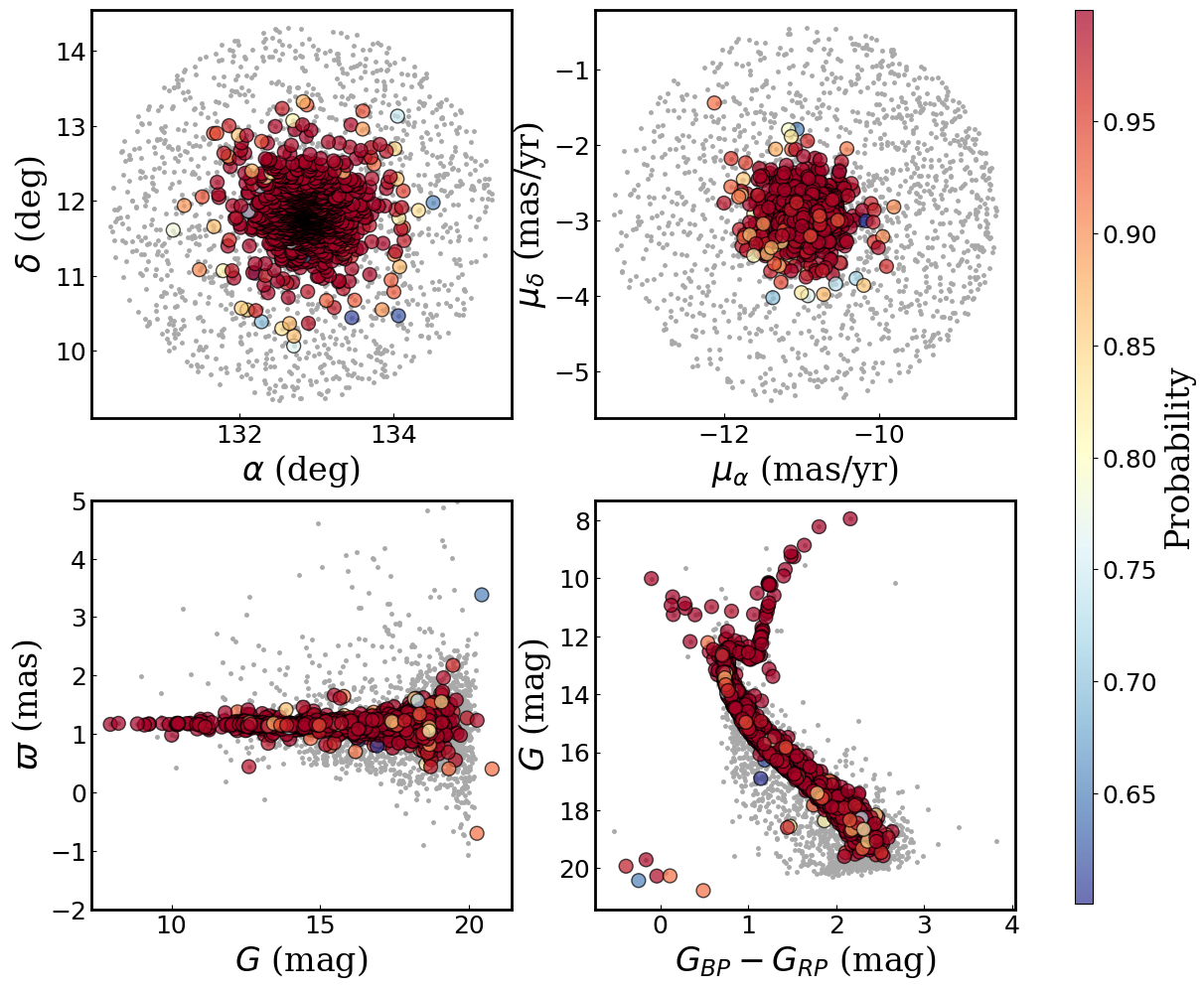}
\caption{Positions in $\alpha$ and $\delta$, VPDs, CMDs, and $\varpi$ as a function of $G$ mag for the cluster M67. In all four panels, the field stars are given by gray dots, and cluster stars are given by the circles with gradient colors. The colorbar present membership probability. \label{fig:M67_result_sample}}
\end{figure}

\indent Following the method for identifying members in \citet{Godoy-Rivera2021} with 3-dimensional(3D) parameter space ($\mu_{\alpha}$, $\mu_{\delta}$, $\varpi$), we consider a 5D parameter space ($\alpha$, $\delta$, $\mu_{\alpha}$, $\mu_{\delta}$, $\varpi$) to identify the members of $Sample$ $Sources$. Assuming that $\mathcal{L}_{c}$ and $\mathcal{L}_{f}$ represent the likelihoods of cluster members and field stars, respectively, the cluster membership probability of a given star $i$ can be expressed as\\
\begin{equation}
P_{c,i} = \frac{n_{c}\mathcal{L}_{c, i}}{n_{c}\mathcal{L}_{i}+n_{f}\mathcal{L}_{i}}
\end{equation}
where $n_{c}$ and $n_{f}$ represent the weight coefficients of cluster members and field stars (normalized by $n_{c}$= 1- $n_{f}$). We assume that the distributions of cluster members and field stars can be approximated as two Gaussian components in a 5D parameter space. For $i$-th sample star, the likelihood of each population (cluster C and field F) can be formed as by 5D multivariate Gaussian
\begin{equation}
\mathcal{L}_{C/F, i} 
= \frac{1}{(2\pi)^{5/2} |\Sigma_{i}|^{1/2}}exp[-\frac{1}{2}(x_{i}-x_{C/F})^{T}\Sigma_{i}^{-1} (x_{i}-x_{C/F})]
\end{equation}
where $\Sigma_{i}$ is the sum of the individual covariance matrix:

$$
\Sigma_{i} =
\left[
    \begin{array}{ccccc}
        \sigma_{\alpha,i}^{2}+ \sigma_{\alpha,P}^{2}
        & \rho_{\alpha\delta,i}\sigma_{\alpha,i}\sigma_{\delta,i}
        & \cdots
        & \rho_{\alpha\mu_{\delta},i}\sigma_{\alpha,i}\sigma_{\varpi, i}\\

        \rho_{\alpha\delta,i}\sigma_{\alpha,i}\sigma_{\delta_{\alpha,i}}
        & \sigma_{\delta,i}^{2} + \sigma_{\delta,P}^{2}
         & \cdots
         & \rho_{\delta\varpi,i}\sigma_{\delta,i}\sigma_{\varpi, i}\\

        \vdots & \vdots & \ddots & \vdots\\

         \rho_{\alpha\mu_{\delta},i}\sigma_{\alpha,i}\sigma_{\varpi, i}
         & \rho_{\delta\varpi,i}\sigma_{\delta,i}\sigma_{\varpi, i}
          & \cdots
          & \sigma_{\varpi,i}^{2} + \sigma_{\varpi,P}^{2}\\
    \end{array}
\right]
$$
and
$$
x_{i}-x_{C/F} =
\left[
    \begin{array}{c}
    \alpha_{i}-\alpha_{C/F}\\
    \delta_{i}-\delta_{C/F}\\
    \mu_{\alpha_{i}}-\mu_{\alpha_{C/F}}\\
    \mu_{\delta_{i}}-\mu_{\delta_{C/F}}\\
    \varpi_{i}-\varpi_{C/F}\\

    \end{array}
\right]
$$
where $\alpha_{C/F}$, $\delta_{C/F}$, $\mu_{\alpha_{C/F}}$, $\mu_{\delta_{C/F}}$, $\varpi_{C/F}$ are the mean values of each population.
The best estimates of cluster and field parameters are determined by solving the above system of equations using the maximum likelihood method.\\
\indent The distribution of cluster membership probabilities for $Sample$ $sources$ in the M67 field is shown in Fig.\ref{fig:M67_prob}. We apply a selection criterion and classify the stars with $P_{c,i}$ $\geq$ 0.9 as high probability members and those with 0.6 $\le$ $P_{c,i}$ $\textless$ 0.9 as moderate probability members. It is worth noting that to mitigate the pollution of field stars as much as possible, we only used high probability members for the research in the following sections, and moderate probability members are also provided in the Table. \ref{tab:Finalcatalog}. We show the positions in $\alpha$ and $\delta$, VPDs, color-magnitude diagrams (CMDs), and $\varpi$ as a function of $G$-band magnitude for the cluster M67 in Fig. \ref{fig:M67_result_sample}.\\

\subsubsection{Members of the clusters}
\begin{figure*}
\gridline{\fig{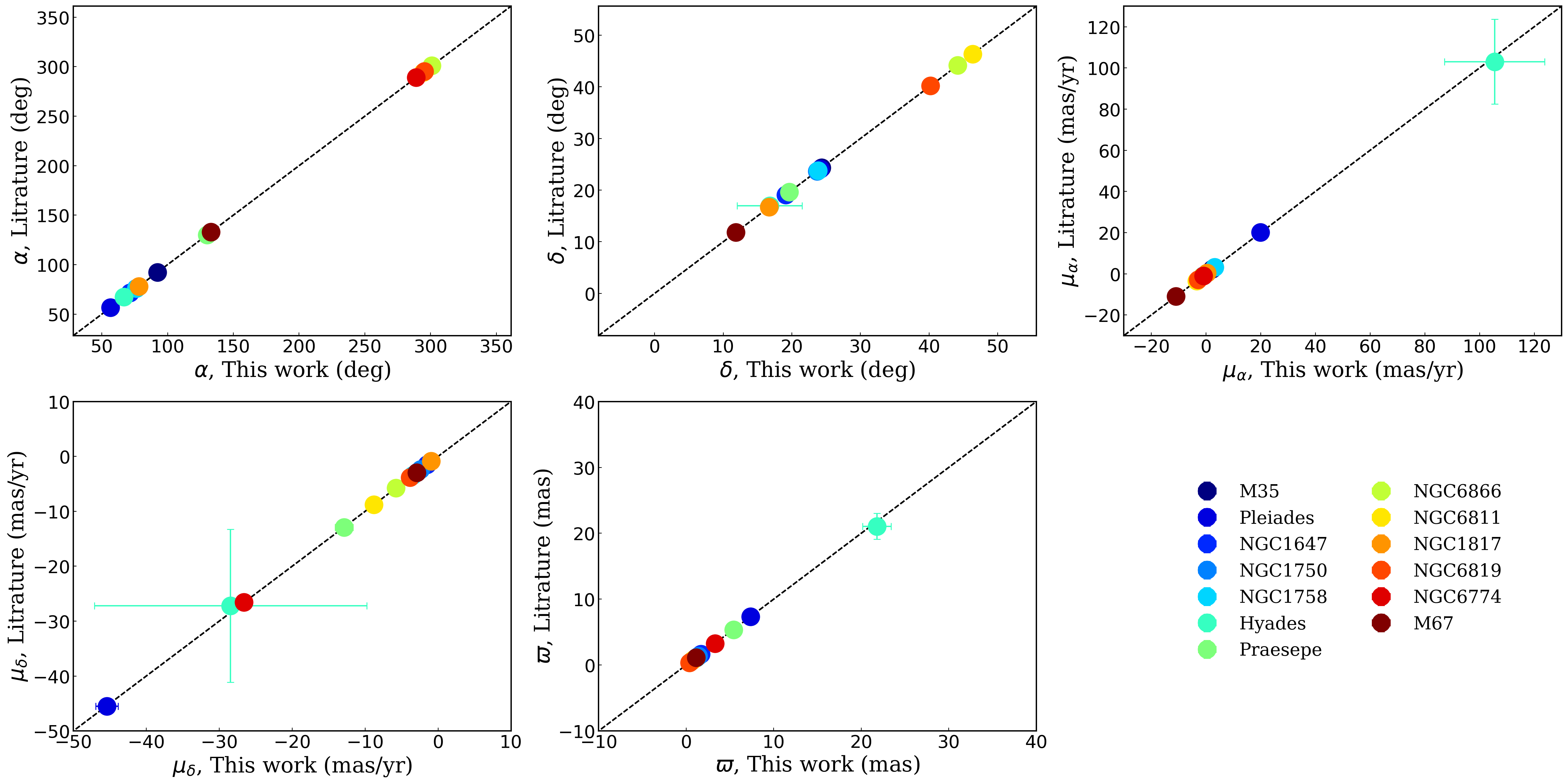}{1\textwidth}{}
          }
\caption{Comparison of our cluster parameters with literature reference values from \citet{Cantat-Gaudin2020A&A...640A...1C}. The black dashed line is 1:1 line.
\label{fig:astro_param}}
\end{figure*}

\begin{figure*}
\gridline{\fig{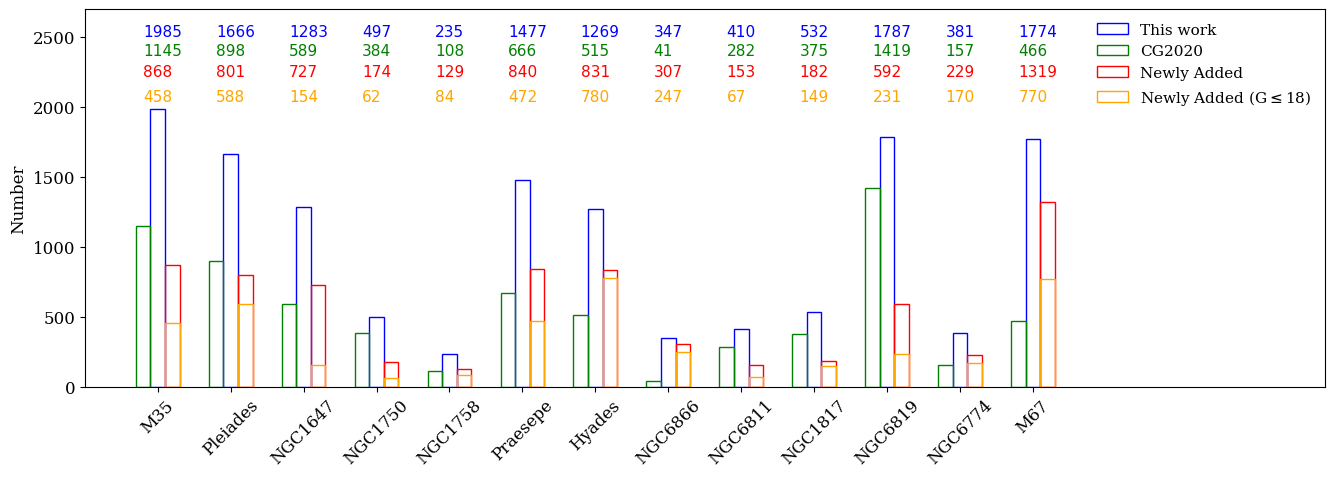}{1\textwidth}{}
          }
\caption{Comparison of our cluster members catalog with \citet{Cantat-Gaudin2020A&A...640A...1C} membership catalog. The blue bars are the numbers of high probability members with $P>0.9$ identified in this work. The green bars are the numbers of probable members ($P>0.7$) published by \citet{Cantat-Gaudin2020A&A...640A...1C}. The red bars are the numbers of new members we added in this work compared to \citet{Cantat-Gaudin2020A&A...640A...1C}. The orange bars are the same as the red bars, but correspond to the new number of members for $G \leq$ 18 mag.}
\label{fig:memb_compare}
\end{figure*}

\indent Young clusters (younger than 20 Myr) are scattered in the ($\mu_{\alpha}$, $\mu_{\delta}$) parameter space due to members tend to have relatively high dispersion of space velocities, making it challenging to identify members with the above method \citep[e.g.][]{Makarov2004MNRAS.352.1199M,Torres2006A&A...460..695T,Kuhn2019ApJ...870...32K,Miret-Roig2022NatAs...6...89M}. Therefore, we apply the above method only to the 13 clusters with ages older than 20 Myr listed in Table \ref{tab:table2}. We derive the mean astrometric parameters ($\alpha$, $\delta$, $\mu_{\alpha}$, $\mu_{\delta}$, $\varpi$) and errors for 13 OCs by fitting a Gaussian to their distributions, and report them in Table \ref{tab:table2}. To ascertain the accuracy of our results, we compare our mean astrometric parameters with those reported by \citet{Cantat-Gaudin2020A&A...640A...1C} in Fig.\ref{fig:astro_param}, which shows that our astrometric parameters are in excellent agreement with them. Fig.\ref{fig:memb_compare} displays membership numbers compared to \citet{Cantat-Gaudin2020A&A...640A...1C} which identified memberships by Gaia DR2. Our method has identified significantly more 100-1300 new members. The increase in the number of members is mainly due to two factors. On the one hand, the G-band magnitude has been extended to 21 mag; on the other hand, the increased accuracy and number of sources observed in Gaia DR3 compared to DR2.\\
\indent For Upper Scorpius and $\rho$ Ophiuchus, the members are determined from the membership catalog of \citet{Miret-Roig2022NatAs...6...89M}, which used proper motions, parallax, and photometry to calculate membership probability. We obtain 2042 members of Upper Scorpius and 735 members of $\rho$ Ophiuchus, respectively. The members of Taurus are determined by membership catalog of \citet{Krolikowski2021AJ....162..110K}, which is currently the most comprehensive census of the Taurus region using Gaia EDR3. We obtain 587 members of Taurus.\\

\section{Cluster Properties} \label{sec:Properties}
\subsection{Spatial structure parameter}
\begin{figure}[ht!]
\plotone{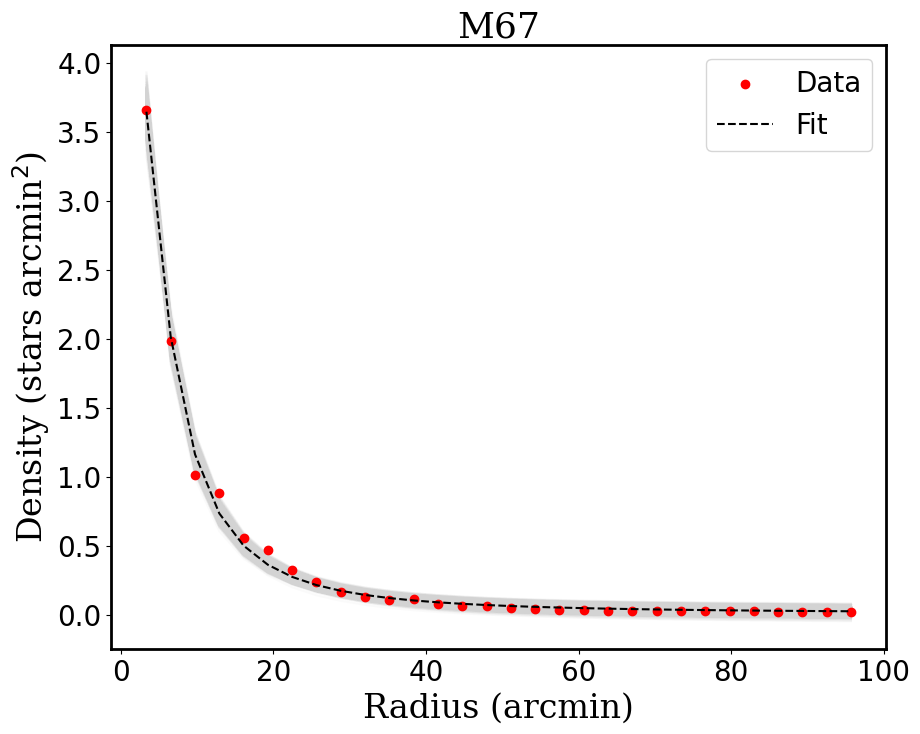}
\plotone{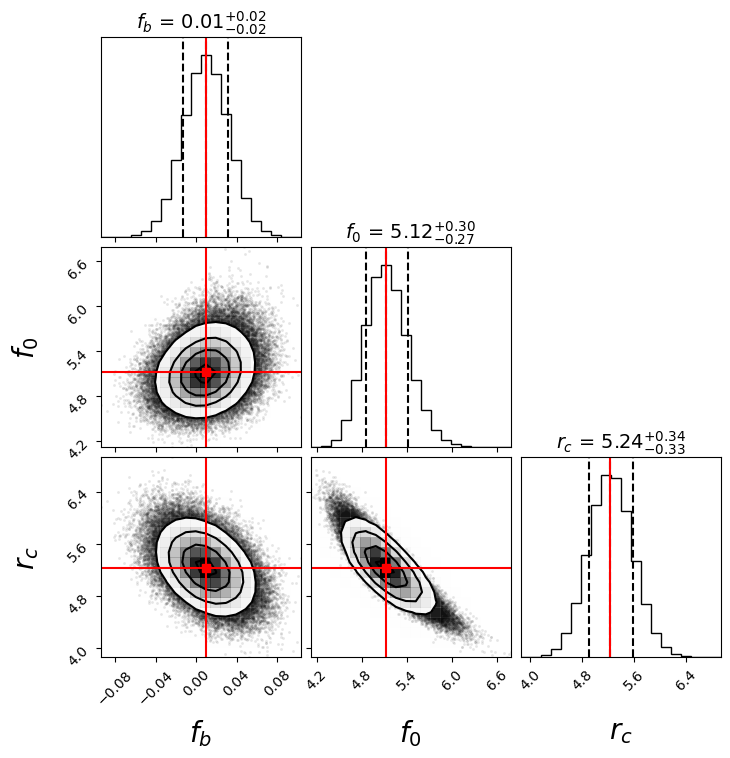}
\caption{The top panel: Results of the King proﬁle ﬁt for cluster M67. The gray lines represent 1000 random sample fits from the MCMC chains, which provide an illustration of the model parameter uncertainties. The black dashed line is best fitting. The bottom panel: Marginalized posterior distribution and uncertainties of model parameters of the King proﬁle ﬁt for the cluster M67. The red solid line on each histogram is the $50^{th}$ percentile of the MCMC samples, and the black dashed lines correspond to the $16^{th}$ and $84^{th}$ percentiles, respectively.\label{fig:M67_RDP}} 
\end{figure}

\indent The radial density profile (RDP) is the traditional tool for investigating the structure of OCs. The number density in each ring by:
\begin{equation}
    \rho _i=\frac{N_i}{\pi \left( r_{i+1}^{2}-r_{i}^{2} \right)}
\end{equation}
where $N_i$ is the number of stars in the $i$th ring with inner and outer radius $r_i$ and $r_{i+1}$. We fit King's surface density profile \citep{King1962AJ.....67..471K} to cluster surface density.
\begin{equation}\label{Kingmodel}
f_r=f_b+\frac{f_0}{1+\left( r/r_c \right) ^2}
\end{equation}
where $f_b$ is the background density, $f_0$ is the central density, $r_c$ is the core radius, and $r$ is the radial distance from the cluster centre. The values of the parameters are calculated using the {\tt emcee}, an open-source Markov Chain Monte Carlo (MCMC) package \citep{Foreman2013PASP..125..306F}. The posterior distribution is set to 1000 walkers, 2000 iterations, and 200 burn-in steps. Fig.\ref{fig:M67_RDP} displays the posterior distributions for each parameter, along with the fitting results for M67. RDP fitting results of other OCs are shown in Appendix \ref{sec:AppendixA}. We consider the $50^{th}$ of the distribution as the best-fit value of the parameters, with uncertainties at the $16^{th}$ and $84 ^{th}$ percentiles, respectively. The calculated values and uncertainties of $r_c$ for each OC are listed in Table \ref{tab:table2}.
\begin{table*}
\scriptsize
\centering
\tabcolsep 0.2truecm
\caption{Astrometric and Spatial structure Parameters.}\label{tab:table2}
\begin{tabular}{cccccccccc}
\hline
\hline
\multicolumn{1}{c}{Cluster name} &
\multicolumn{1}{c}{$\alpha (J2000)$}&
\multicolumn{1}{c}{$\delta (J2000)$}&
\multicolumn{1}{c}{$\mu_{\alpha}$}&
\multicolumn{1}{c}{$\mu_{\delta}$} &
\multicolumn{1}{c}{$\varpi$}&
\multicolumn{1}{c}{$R_{c}$}&
\multicolumn{1}{c}{Members}&
\multicolumn{1}{c}{Members}&
\\
\multicolumn{1}{c}{} &
\multicolumn{1}{c}{(deg)}&
\multicolumn{1}{c}{(deg)}&
\multicolumn{1}{c}{(mas/yr)}&
\multicolumn{1}{c}{(mas/yr)} &
\multicolumn{1}{c}{(mas)}&
\multicolumn{1}{c}{(arcmin)}&
\multicolumn{1}{c}{$P>0.9$}&
\multicolumn{1}{c}{$P>0.6$}&
\\
\hline
M35 & 92.26 $\pm$ 0.35 & 24.34 $\pm$ 0.37 & 2.27 $\pm$ 0.28 & -2.9 $\pm$ 0.26 & 1.19 $\pm$ 0.15            & 9.31   $\pm$ 0.08    & 1985 & 2484   \\       
Pleiades & 56.6 $\pm$ 1.58 & 24.08 $\pm$ 1.54 & 19.96 $\pm$ 1.31 & -45.47 $\pm$ 1.81 & 7.39 $\pm$ 0.33     & 35.88  $\pm$ 1.08    & 1667 & 1677    \\      
NGC 1647 & 71.47 $\pm$ 0.47 & 19.08 $\pm$ 0.4 & -1.12 $\pm$ 0.41 & -1.55 $\pm$ 0.34 & 1.77 $\pm$ 0.29       & 12.32  $\pm$ 0.35    & 1283 & 1385   \\      
NGC 1750 & 75.92 $\pm$ 0.36 & 23.66 $\pm$ 0.3 & -1.01 $\pm$ 0.23 & -2.4 $\pm$ 0.18 & 1.43 $\pm$ 0.16        & 9.56   $\pm$ 0.44    & 497 & 614     \\      
NGC 1758 & 76.19 $\pm$ 0.13 & 23.81 $\pm$ 0.11 & 3.14 $\pm$ 0.16 & -3.49 $\pm$ 0.13 & 1.16 $\pm$ 0.13       & 3.04   $\pm$ 0.03    & 235 & 275      \\     
Hyades & 65.93 $\pm$ 10.28 & 15.73 $\pm$ 10.49 & 100.13 $\pm$ 37.39 & -31.86 $\pm$ 32.7 & 21.19 $\pm$ 2.74 & 33.21  $\pm$ 1.08    & 1269 & 1435    \\      
Praesepe & 130.1 $\pm$ 1.87 & 19.61 $\pm$ 1.82 & -36.25 $\pm$ 2.51 & -12.91 $\pm$ 1.85 & 5.42 $\pm$ 0.35   & 126.24 $\pm$ 11.25   & 1477 & 1522    \\      
NGC 6866 & 300.99 $\pm$ 0.18 & 44.19 $\pm$ 0.16 & -1.36 $\pm$ 0.11 & -5.77 $\pm$ 0.1 & 0.7 $\pm$ 0.06       & 2.1    $\pm$ 0.01    & 347 & 487       \\    
NGC 6811 & 294.34 $\pm$ 0.26 & 46.36 $\pm$ 0.18 & -3.35 $\pm$ 0.14 & -8.81 $\pm$ 0.14 & 0.91 $\pm$ 0.11     & 4.02   $\pm$ 0.05    & 410 & 494       \\    
NGC 1817 & 78.14 $\pm$ 0.16 & 16.7 $\pm$ 0.15 & 0.42 $\pm$ 0.09 & -0.93 $\pm$ 0.08 & 0.58 $\pm$ 0.07        & 3.16   $\pm$ 0.05    & 532 & 773       \\    
NGC 6819 & 295.33 $\pm$ 0.09 & 40.19 $\pm$ 0.07 & -2.9 $\pm$ 0.11 & -3.87 $\pm$ 0.12 & 0.38 $\pm$ 0.06      & 2.33   $\pm$ 0.05    & 1787 & 2503      \\   
NGC 6774 & 289.11 $\pm$ 1.18 & -16.37 $\pm$ 1.94 & -0.91 $\pm$ 0.77 & -26.82 $\pm$ 1.35 & 3.34 $\pm$ 0.29   & 14.4   $\pm$ 0.26    & 381 & 586        \\   
M67 & 132.85 $\pm$ 0.34 & 11.83 $\pm$ 0.33 & -10.96 $\pm$ 0.3 & -2.9 $\pm$ 0.26 & 1.16 $\pm$ 0.21          & 5.41   $\pm$ 0.21    & 1774 & 1844      \\    
\hline
\end{tabular}
\end{table*}

\subsection{The Metallicity, Distance modulus, Reddening and Age} \label{sec:age}
\begin{figure*}
\centering
\includegraphics[width=1\textwidth]{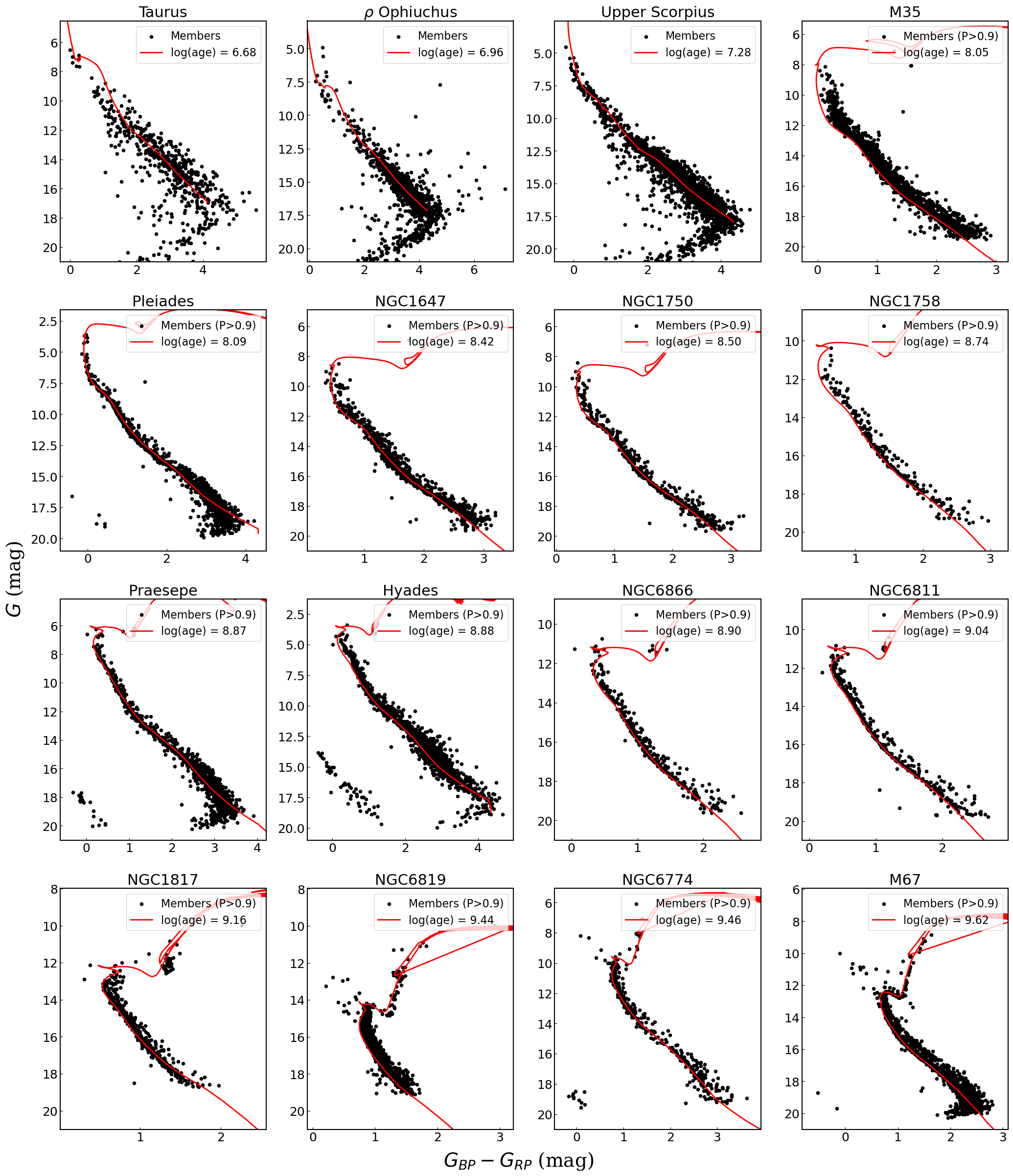}
\caption{CMDs and isochrone fittings of 16 OCs. The black dots are the members with $P > 0.9$.
\label{fig:isofit}}
\end{figure*}

\indent Thanks to the release of large-scale spectroscopic surveys such as LAMOST\citep{Cui2012RAA....12.1197C,Zhao2012RAA....12..723Z,Deng2012RAA....12..735D}, GALAH\citep{Silva2015MNRAS.449.2604D}, which provide us with accurate metallicity. We adopt [Fe/H] of 12 OCs from \citet{Fu2022arXiv220709121F} who reported metallicity with LAMOST DR8 \footnote{\url{http://www.lamost.org/dr8/}} spectra. For NGC 6774, we use the [Fe/H] = 0.12 of \citet{Donor2020AJ....159..199D} as its metallicity. For Taurus, we cross-match the reliable member stars with LAMOST DR8 catalog and obtain [Fe/H] for 41 members. Similarly, for Upper Scorpius and $\rho$ Ophiuchus, we cross-match the reliable member stars with GALAH DR3\footnote{\url{https://www.galah-survey.org/dr3/overview/}} catalog and obtain [Fe/H] for 85 and 16 members, respectively. Then, we follow the method of \citet{Fu2022arXiv220709121F} to calculate the [Fe/H] of these three OCs. The metallicities of 16 OCs are listed in Table \ref{tab:isofitting}.\\  
\indent To determine the ages of OCs, we calculate the distance between the member star and the closest point of the isochrone by the nearest neighbour  algorithm \citep{2019ApJS..245...32L}. The mean distance was calculated with the following formula:
\begin{equation}
\overline{d^2}=\sum_{k=1}^n{|x_k-x_{k,nn}|^2/n}
\end{equation}
where $x_k$=$\left[ G_k+\Delta G,\ \left( G_{BP}-G_{RP} \right) _k+\Delta \left( G_{BP}-G_{RP} \right) \right] $ is the position of the $k$th star in the CMD. Here, $\Delta G$ ($m-M$) is the distance modulus in $G$ magnitude, and $\Delta \left( G_{BP}-G_{RP} \right)$ is the color excess $E(BP-RP)$. $x_{k, nn}$ is the position of the isochrones, which are closest to the corresponding $k$th star. 
A series of isochrones from the PARSEC database\footnote{\url{http://stev.oapd.inaf.it/cgi-bin/cmd}} \citep{Bressan2012MNRAS.427..127B} is generated with fixed metallicity, covering a range of log(Age/yr) from 6.0 to 10.0 with a step size of 0.01. We fit the CMD and find a best isochrone for each OC. Three parameters: age, $m-M$, $E(BP-RP)$ are obtained when $\overline{d^2}$ is the minimum. The isochrone fitting parameters are shown in Table \ref{tab:isofitting} and the results of fitting the CMDs of OCs are shown in Fig \ref{fig:isofit}. To verify our results, we compare the ages obtained in this work with those reported in \citet{Bossini2019A&A...623A.108B} and show that they are in good agreement.\\
\begin{table}[h!t]
\scriptsize
\centering
\tabcolsep 0.2truecm
\caption{Isochrone fitting parameters of 16 OCs.}\label{tab:isofitting}
\begin{tabular}{lcccccccccc}
\hline
\hline
\multicolumn{1}{c}{Cluster} &
\multicolumn{1}{c}{Age}&
\multicolumn{1}{c}{[Fe/H]}&
\multicolumn{1}{c}{$m-M$}&
\multicolumn{1}{c}{$E(BP-RP)$}&
\\
& (Myr) & (dex) & (mag) &(mag) \\
\hline
Taurus & 4 & -0.025$^{c}$ & 6.002 & 0.32\\
$\rho$ Ophiuchus & 9 & -0.413$^{d}$ & 5.577 & 0.361\\
Upper Scorpius & 19 & -0.531$^{d}$ & 5.627 & 0.277\\
M35 & 112 & -0.054$^{a}$ & 9.995 & 0.226\\
Pleiades & 123 & 0.096$^{a}$ & 5.684 & 0.126\\
NGC 1647 & 263 & 0.003$^{a}$ & 9.796 & 0.59\\
NGC 1750 & 316 & -0.041$^{a}$ & 9.999 & 0.479\\
NGC 1758 & 549 & -0.011$^{a}$ & 11.429 & 0.695\\
Praesepe & 741 & 0.248$^{a}$ & 6.114 & 0.003\\
Hyades & 758 & 0.279$^{a}$ & 3.497 & 0.032\\
NGC 6866 & 794 & 0.0153$^{a}$ & 11.131 & 0.244\\
NGC 6811 & 1096 & -0.073$^{a}$ & 10.359 & 0.121\\
NGC 1817 & 1445 & -0.215$^{a}$ & 11.217 & 0.261\\
NGC 6819 & 2754 & 0.046$^{a}$ & 12.082 & 0.174\\
NGC 6774 & 2884 & 0.12$^{b}$ & 7.57 & 0.176\\
M67 & 4168 & -0.018$^{a}$ & 9.614 & 0.054\\
\hline
\\
\end{tabular}
\tablecomments{$^a$ values of collected from \cite{Fu2022arXiv220709121F};
$^b$ values of collected from \cite{Donor2020AJ....159..199D};
$^c$ median values of LAMOST DR8 catalog;
$^d$ median values of GALAH DR3 catalog.}
\end{table}

\section{Binary and blue straggler star candidates}\label{sec:binary}
In the Fig \ref{fig:isofit}, it is apparent that some members display higher luminosity and occupy the brighter and redder region of the main sequence (MS) single stars in the CMDs. These members are primarily binaries. Additionally, blue straggler stars (BSSs), which are brighter and bluer than the main sequence turn-off (MSTO) in CMDs of OCs, are present in older OCs such as NGC 1817, NGC 6819, NGC 6774, and M67. Based on available data, we use different methods to identify the astrometric, photometric, and spectroscopic binary candidates and BSSs.

\subsection{Astrometric Binaries}
\indent The Renormalized Unit Weight Error (RUWE) parameter in Gaia is the square root of the reduced $\chi^2$ statistic for the astrometric solution\citep{Lindegren2018A&A...616A...2L,Pearce2020ApJ...894..115P}. The RUWE is expected to be about 1.0 when the source is a single star. For unresolved binaries or tertiary companion stars, the optical center motion and mass decouple will lead to poor performance of single-source astrometric models \citep{Belokurov2020MNRAS.496.1922B,Stassun2021ApJ...907L..33S}. A high RUWE value ($\geq$1.0) may indicate that a source is non-single \citep{Belokurov2020MNRAS.496.1922B}. We adopt a criterion of RUWE $\geq$ 1.4 to detect multi-object systems\citep[e.g.][]{Jorissen2019MmSAI..90..395J,Belokurov2020MNRAS.496.1922B}, and flag 1,851 stars as astrometric binaries (marked as ${\tt N_{ruwe}}$ listed in Table \ref{tab:BF}).\\

\subsection{Photometric Binaries}
Photometric binaries are commonly identified as stars that appear brighter and redder than their counterparts on the MS in a CMD. Therefore, the binary properties can be deduced from the statistical analysis based on CMD. One common approach to identify binaries in the CMD is dividing the CMD into single and binary regions by mass ratio ($q$) \citep[see, e.g., ][]{Milone2012A&A...540A..16M, Li2013MNRAS.436.1497L, Li2020ApJ...901...49L}.\\
\indent We use interpolated PARSEC isochrones to calculate the isochrones for various mass ratios. A binary system with $q$ =1 (equal-mass binary) will appear -2.5log(2) $\approx$ 0.752 mag brighter than the MS star. Conversely, binary with small $q$ values becomes indistinguishable from MS star. When the $q$ is neither equal to 1 nor close to 0, the binaries will gather on the reddish side of the MS in the CMD, but below the equal-mass binary sequence. Thus, we can distinguish binaries from single stars by setting a limit on the mass ratio ($q_{min}$). We set $q_{min}$ = 0.6 following \citet{Jadhav2021AJ....162..264J}.  Stars with a magnitude brighter than the isochrone of $q_{min}$ in the same color are flagged as photometric binary candidates. We identify 2,716 stars as photometric binary candidates from 13 OCs, and the numbers for each cluster are listed in Table \ref{tab:BF} (marked as ${\tt N^{0.6}_{phot.}}$). For Taurus, Upper Scorpius and $\rho$ Ophiuchus, many stars have not evolved to the main sequence, so this feature is not displayed on CMD. \\

\subsection{Spectroscopic Binaries}
\indent Spectroscopic binaries (SBs) are identified by observing the periodic variation of radial velocity (RV) with time. Recent studies on SB have been conducted in large-scale spectroscopic surveys, such as APPOGEE\citep{Kounkel2021AJ....162..184K,Price-Whelan2020ApJ...895....2P}, LAMOST\citep{Kovalev2022MNRAS.517..356K,Tian2020ApJS..249...22T}, GLAHA\citep{Traven2020A&A...638A.145T}, RAVE\citep{Birko2019AJ....158..155B}. In addition, there are also studies of SBs for single clusters, such as M35 \citep{Leiner2015AJ....150...10L}, Hyades\citep{Bender2008ApJ...689..416B,Griffin2012JApA...33...29G}, NGC 6819 \citep{Milliman2014AJ....148...38M}, M67\citep{Geller2021AJ....161..190G}. We cross-match our high probability members with these studies to determine SBs. We identify 692 SBs among our members, and the numbers for each cluster are listed in Table \ref{tab:BF} (marked as ${\tt N_{SB}}$). The information of binary candidates with different symbols in Fig. \ref{fig:binary}.


\subsection{Binary Fraction}
\begin{figure}[ht!]
\plotone{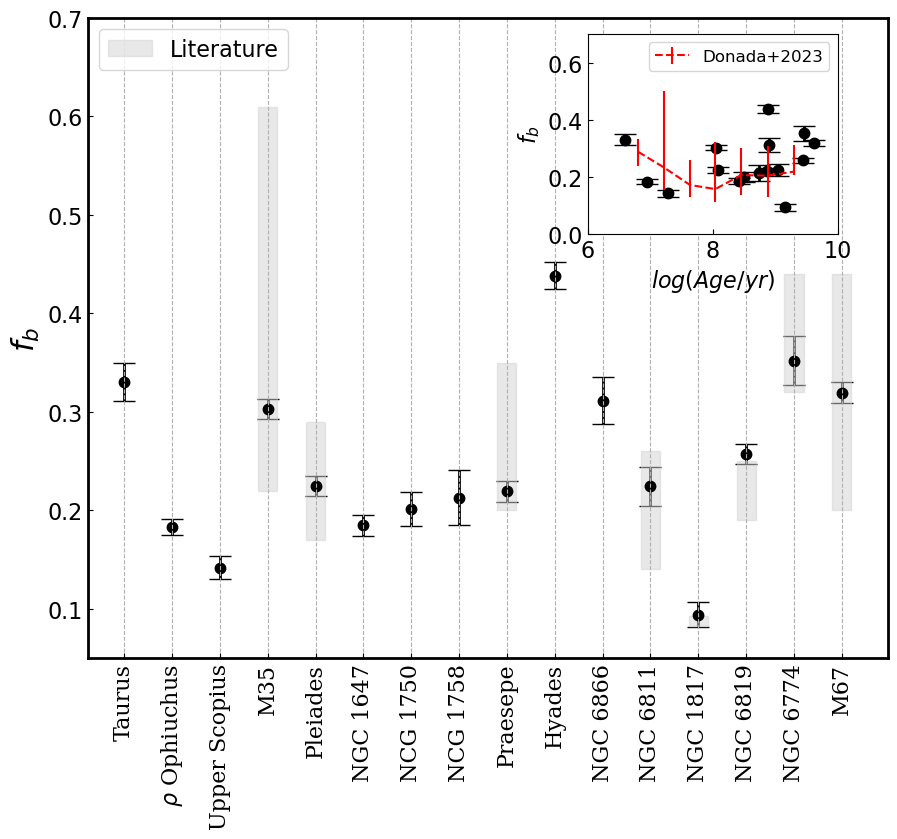}
\caption{ Distribution of BF for 16 clusters. The clusters are arranged in increasing order of age. The black dots are BFs obtained in this work, and gray rectangles indicate the range of BFs obtained by previously published work. Subplot is the BF of systems as a function of logarithmic age. The red dashed line is the result of \citet{Donada2023arXiv230111061D}.\label{fig:BF}}.
\end{figure}

\indent Binary Fraction (BF) in a cluster is critical to understanding the influence of binaries on the properties and dynamical evolution of the host cluster. The BF in this work is calculated as 
\begin{equation}
    f_{b} =\frac{N_{b,tot}}{N_s+N_{b,tot}}
\end{equation}
where $N_s$ is the number of single members, $N_{b,tot}$ is the numbers of binary members. To estimate the uncertainties in BF, we utilize the bootstrap method described by \cite{Jadhav2021AJ....162..264J}. We measure BF for each cluster using 1000 random samples. The uncertainty of BF was determined as the standard deviation of the 1000 measurements obtained. The BFs and uncertainties of clusters are listed in Table \ref{tab:BF}. \\
\indent Table \ref{tab:BF} illustrates the BFs of the OCs vary from 9\% to 44\%. We compare the BF values in this work with previous studies, the comparison results are shown in Table \ref{tab:BF} and Fig.\ref{fig:BF}. The BFs of the seven OC (M35, Pleiades, Preasepe, NGC 6811, NGC 1817, NGC 6774, M67) estimated in this work are consistent with the results in previous studies, whereas NGC 6819 is slightly lower than that of the literature. Our estimated BF for NGC 6819 is $25.72 \% \pm 4.34\%$, slightly higher than the results ($22 \% \pm 3\%$) obtained by \citet{Milliman2014AJ....148...38M} based on 93 spectroscopic binaries. The subplot in Fig.\ref{fig:BF} shows the relationship between BF and age, the trend is consistent with that of \cite{Donada2023arXiv230111061D} (red dashed line). \\

\begin{table}[]
\scriptsize
\centering
\tabcolsep 0.07truecm
\caption{Binary Fractions of 16 OCs.}\label{tab:BF}
\begin{tabular}{lccccccc}
\hline
\hline
 & \multicolumn{5}{c}{This Work}                               & Reference  \\ \cline{2-6} 
Clusters    & N$_{ruwe}$ & N$^{0.6}_{phot.}$   & N$_{SB}$ & N$_{b,tot}$   & $f_b$    & $f_b$       \\
         &      &       &  &  & \%    & \%    \\ \hline

Taurus          &  175    &        &    29   &    192 &   33.04$\pm$1.89 \\                           
$\rho$ Ophiuchus&  130    &        &    7    &    135 &   14.19$\pm$1.13 \\                           
Upper Scorpius  &  393    &        &    33   &    413 &   18.33$\pm$0.79 \\                           
M35             &  121    &   523  &    74   &    601 &  30.28$\pm$4.62    &    $50\pm8^a$      \\    
                &         &        &         &        &                    &    $23\pm1^b$      \\    
                &         &        &         &        &                    &    $24\pm3^c$      \\    
                &         &        &         &        &                    &    $61^d$   \\           
Pleiades        &  204    &   238  &    96   &    374 &  22.35$\pm$4.17    &    $20\pm3^a $      \\   
                &         &        &         &        &                    &    $23^{+6 d}_{-5}$ \\   
                &         &        &         &        &                    &    $14\pm2^b$      \\    
NGC 1647        &  83     &   178  &    1    &    236 &  18.66$\pm$3.95    &                     \\   
NGC 1750        &  31     &   76   &    0    &    98  &  20.14$\pm$4.01    &                      \\  
NGC 1758        &  5      &   48   &    0    &    50  &  21.25$\pm$4.07    &                       \\ 
Praesepe        &  182    &   208  &    35   &    321 &  22.0 $\pm$4.13    &     $35^e $           \\ 
                &         &        &         &        &                    &     $25.3\pm5.4^f$   \\  
Hyades          &  315    &   413  &    57   &    556 &  43.83$\pm$4.97    &                       \\ 
NGC 6866        &  13     &   102  &    1    &    108 &  31.1 $\pm$4.69    &                       \\ 
NGC 6811        &  15     &   72   &    13   &    91  &  22.49$\pm$4.2     &     $20\pm6^a $       \\ 
NGC 1817        &  10     &   40   &    0    &    48  &  9.44 $\pm$2.92    &     $8.3^g $          \\ 
NGC 6819        &  31     &   390  &    73   &    460 &  25.72$\pm$4.34    &     $22\pm3^h $       \\ 
NGC 6774        &  71     &   83   &    12   &    132 &  35.28$\pm$4.76    &     $38\pm6^b $       \\ 
M67             &  72     &   345  &    261  &    566 &  31.98$\pm$4.63    &     $22\pm2^b $       \\ 
\hline 
total          &  1851	  &  2716  &    692	  &  4381  \\             
\hline
\end{tabular}
\tablecomments{a: \citet{Niu2020ApJ...903...93N}; b: \citet{Jadhav2021AJ....162..264J}; c: \citet{Leiner2015AJ....150...10L}; d: \citet{Thompson2021AJ....161..160T}; e: \citet{Pinfield2003MNRAS.342.1241P}; f: \citet{Borucki2010Sci...327..977B} f: \citet{Khalaj2013MNRAS.434.3236K}; g: \citet{Cordoni2018ApJ...869..139C}; h:\citet{Milliman2014AJ....148...38M}; i:\citet{Geller2021AJ....161..190G}  }
\end{table}

\subsection{Blue Straggler Stars}
\begin{figure*}
\centering
\includegraphics[height=4.3cm,width=4.1cm]{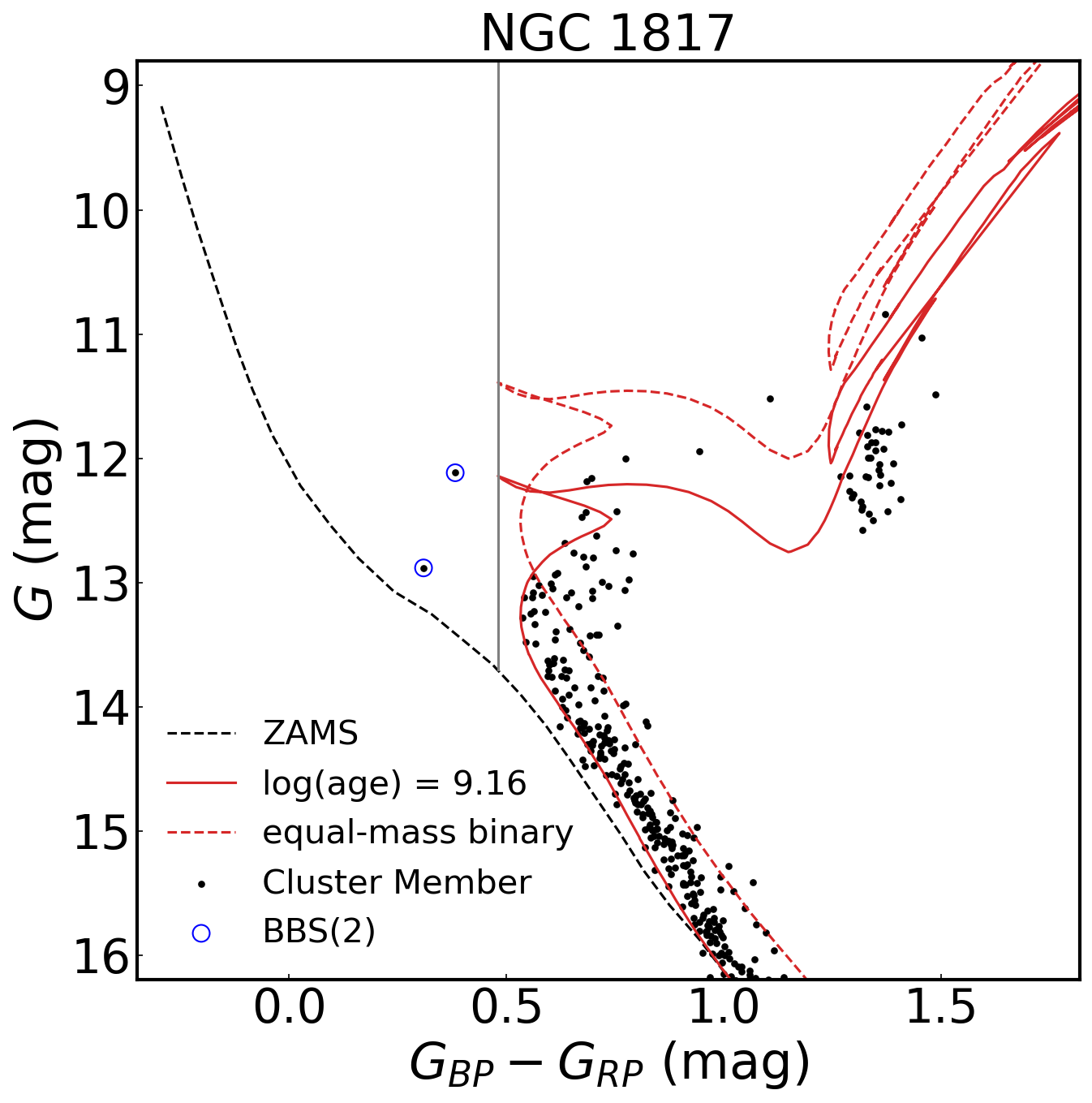}
\includegraphics[height=4.3cm,width=4.1cm]{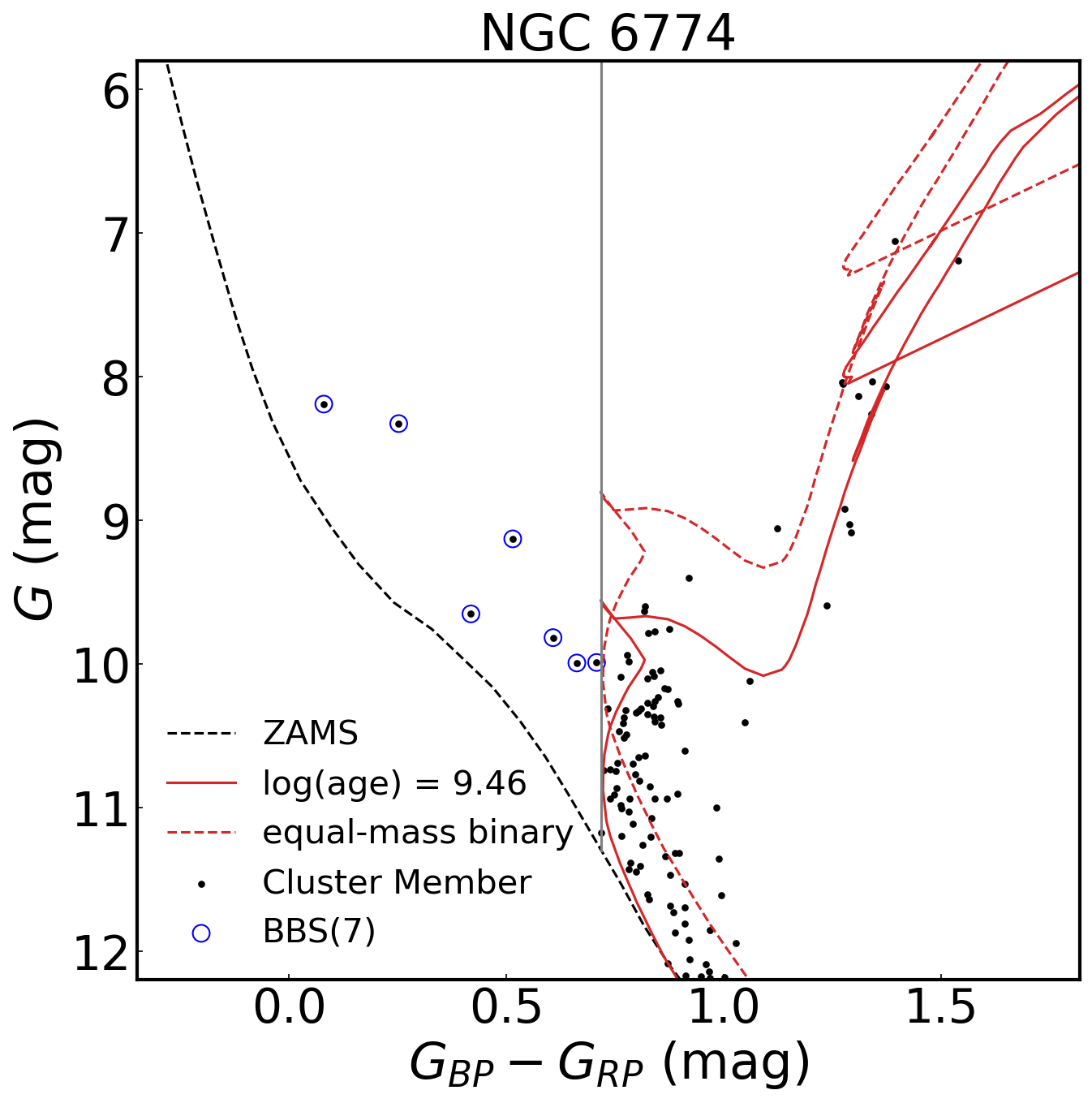}
\includegraphics[height=4.3cm,width=4.1cm]{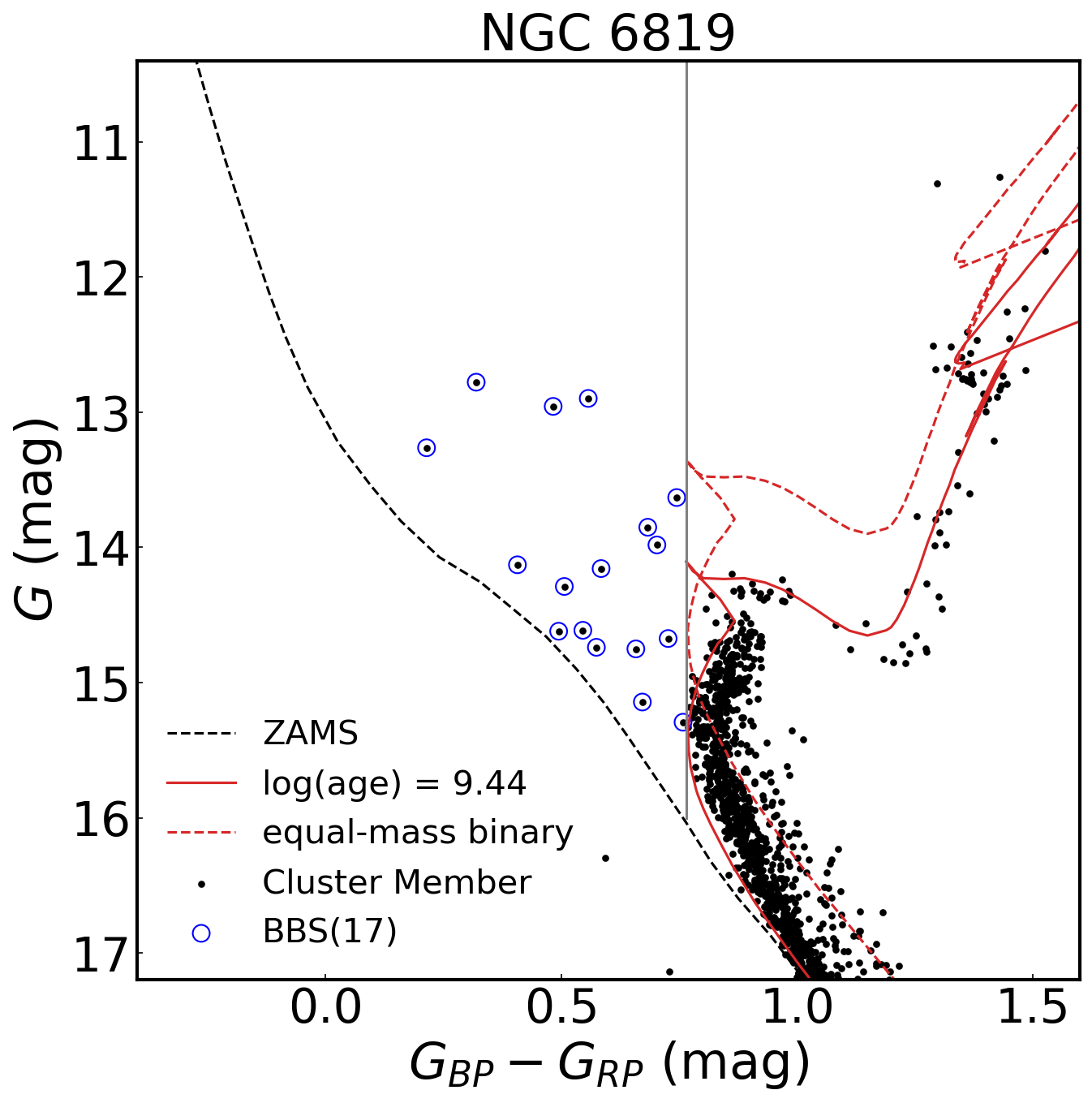}
\includegraphics[height=4.3cm,width=4.1cm]{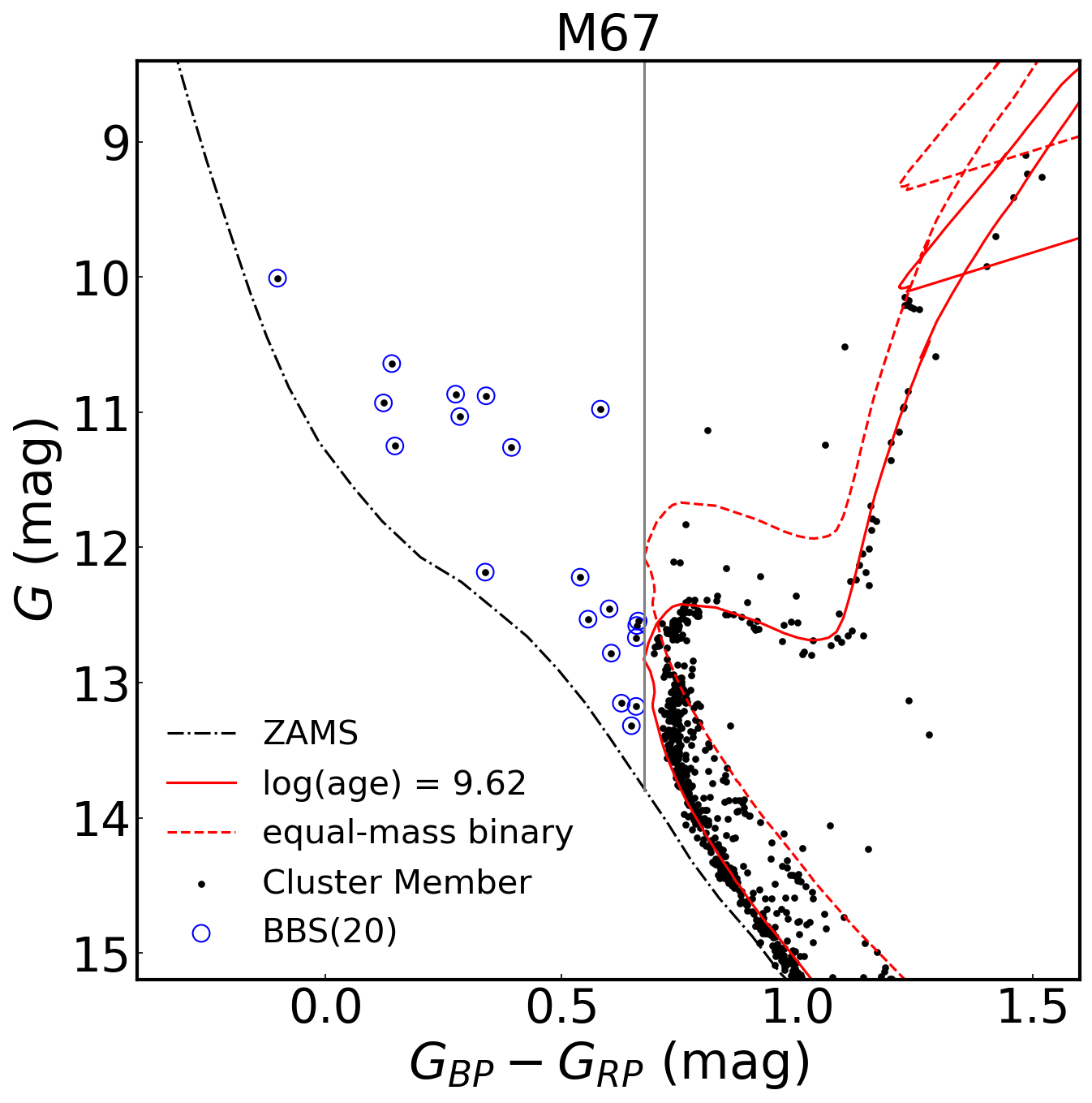}
\caption{CMDs of the NGC 1817, NGC 6774, NGC 6819 and M67. Black filled circles are cluster members with $P$ $\geq$ 0.9. Blue circles are BSS candidates. The red solid line and red dotted line represent the PARSEC isochrones and binary sequences, resepectively. The gray solid line is the red limit by the MSTO colour and the binary sequence as red boundary. The black dashed line is ZAMS as blue limit.
\label{fig:BSS}}
\end{figure*}

\indent To identify BSS candidates, we refer to the definition of \citet{2021A&A...650A..67R} for the region on the CMD where the BSS candidates are located. Fig.\ref{fig:BSS} shows the limited region to select BSS. The constraints are notated in Fig.\ref{fig:BSS}, including the Zero Age Main-Sequence (ZAMS), MSTO color, the lower limit by the mag where the sample stars locate separately from the ZAMS, and the equal-mass binary isochrone that could exclude binaries.
We detect 2, 8, 17 and 22 BSS candidates in NGC 1817, NGC 6774, NGC 6819 and M67, respectively. In Fig.\ref{fig:BSS}, the BSSs of these four clusters are shown in the CMDs by the notation of blue circles.

\section{Rotation} \label{sec:Rotation}
\subsection{Determining the rotation periods}
\begin{figure*}
\gridline{\fig{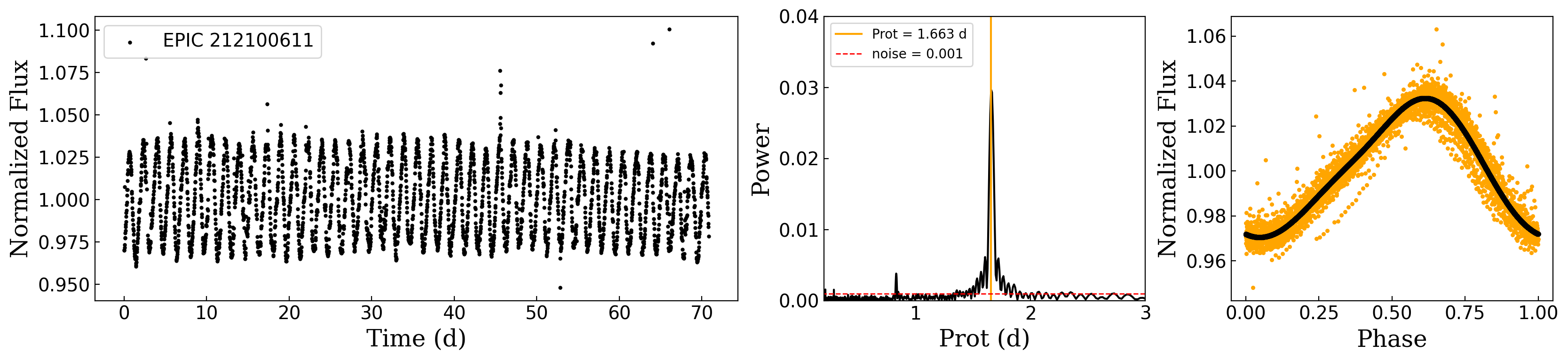}{1\textwidth}{}
          }
\caption{An example for a star which we measure a confident $P_{\rm rot}$, EPIC 211033487. $Left$: Light curve for this star. $Middle$: Lomb-Scargle periodogram, the red dashed line is white noise, which calculated by root nean square of powers. $Right$: Phase-folded light curve using the $P_{\rm rot}$.
\label{fig:prot_example}}
\end{figure*}

\begin{figure}[ht!]
\plotone{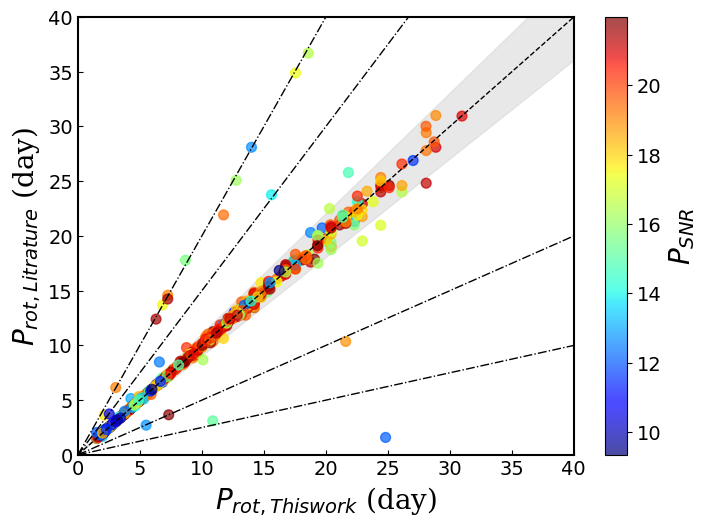}
\caption{Comparison between $P_{\rm rot}$ measured for MS stars in this work with with a previously measured $P_{\rm rot}$ \citep{Douglas2016,Douglas2019,Rebull2016,Rebull2017, Curtis2019ApJ...879...49C, Curtis2020ApJ...904..140C, Rampalli2021, Godoy-Rivera2021}. The dotted lines in the figure depict the relation $P_{\rm rot, thiswork}$ = $\alpha \times$ $P_{\rm rot, literature}$, $\alpha$ = [0.25, 0.5, 1, 1.5, 2]. The gray band corresponds to difference $\leq$ 10\% from 1:1 match. The colorbar presents SNR of $P_{\rm rot}$. 
\label{fig:prot_ref}}
\end{figure}

\indent We cross-match the high probability members ($P \geq$ 0.9) identified in Section \ref{sec:Membership} with $Kepler$\footnote{\url{https://archive.stsci.edu/kepler/data_search/search.php}} and $K2$\footnote{\url{https://archive.stsci.edu/k2/epic/search.php}} mission archives, resulting in the light curves of 827 stars from the Kepler mission, and 4,785 stars from the K2 mission. We use the Lomb-Scargle periodogram with the pre-search data conditioning simple aperture photometry \citep[PDCSAP;][]{Stumpe2012PASP..124..985S,Smith2012PASP..124.1000S} light curves to measure rotation periods ($P_{\rm rot}$). The computed Lomb-Scargle periodogram power ranges from 0.5 to 45 days (approximately half the length of the campaign).\\
\indent We consider the highest peak as the default $P_{\rm rot}$. To evaluate the significance of the peaks, we calculate the signal-to-noise ratio (SNR) of the peaks, which is determined by the following equation:
\begin{equation}
    SNR\ =\ \frac{LS-\overline{LS}}{noise}
\end{equation}
where 
\begin{equation}
LS\ =\ \frac{\chi _{0}^{2}-\chi \left( f \right)^2}{\chi _{0}^{2}}
\end{equation}
$\chi _{0}^{2}$ is the value of $\chi ^2$ about the weighted mean and $\chi \left( f \right)^2$ is the value of $\chi ^2$ about the best-fit sinusoidal with frequency $f$, and $noise$ is the mean root square of $LS$. We set the threshold of the SNR to 7 below periodogram peaks are insignificant according to \cite{Soares2020}. An example of a star in which we measure a confident $P_{\rm rot}$ is shown in Fig.\ref{fig:prot_example}. Following the method of \citet{Serna2021ApJ...923..177S}, we use a bootstrap method to estimate $P_{\rm rot}$ uncertainty. Each light curve was perturbed within the flux uncertainties to generate 300 simulated light curves, and the $P_{\rm rot}$ values were recalculated based on the simulated light curves. Finally, the $P_{\rm rot}$ uncertainties were estimated as the standard deviation of the 300 measurements. To obtain high-confidence rotators, we remove rotators with $P_{\rm rot}$ uncertainty greater than 5 days.\\
\indent Some of members show multiple significant periodic peaks, and secondary peaks can generally be divided into two categories: (1) harmonics of true period due to symmetrical spot conﬁguration and/or spot evolution on the stellar surface \citep{McQuillan2013MNRAS.432.1203M}; (2) true secondary period. For harmonics, we double the aliased $P_{\rm rot}$ and use the corrected value as the final period. Stars are considered to have multiple period when the two peaks are separated at least 50 \% of the primary period.\\ 
\indent For a given star sample, we compute Lomb-Scargle periodograms for all available quarters or campaigns, and report the median and standard deviation values for the resulting $P_{\rm rot}$ and the uncertainties. We measure 4,304 rotators in our sample with the Kepler/K2 lightcurves. We further collect 1,163 rotators based on ground-based data, and finally obtain 5,467 rotators. We compare our $P_{\rm rot}$ for about 2,000 MS stars with a previously measured $P_{\rm rot}$, as illustrated in Fig. \ref{fig:prot_ref}. We find that 98\% of measurements are in agreement to within 10\% margin, with approximately 2\% exhibiting a harmonic period offset by a factor of two. This high level of agreement provides strong support for the validity of our methods and the accuracy of our $P_{\rm rot}$ measurements.\\
\indent In order to exclude some variable stars such as $\delta$ Scuti variables, $\gamma$ Doradus variables, $\beta$ Cephei variables, we restrict our sample to MS cool stars with an effective temperature range of 2800 to 6800 K (corresponding to $(G_{BP}- G_{RP})_0$ values from 0.5 to 4 mag) that have undergone wind-driven rotational evolution and are suitable for gyrochronology. These variable stars lie outside this effective temperature range. In addition, we screen out 66 eclipsing binaries by examining the shape of light curves. The full sample of member stars of 16 OCs with information on membership, binary and rotation is presented in Table \ref{tab:Finalcatalog}.

\begin{table*}
\scriptsize
\centering
\tabcolsep 0.5truecm
\caption{Column overview of the final membership catalog.}\label{tab:Finalcatalog}
\begin{tabular}{lcll}
\hline
\hline
\multicolumn{1}{c}{Patameter} &
\multicolumn{1}{c}{Format}&
\multicolumn{1}{c}{Units}&
\multicolumn{1}{c}{Description}
\\
\hline
Cluster                 & string    &    -     &  Cluster name \\
Source              & long    &    -       &   Gaia DR3 ID                                             \\
RAdeg                     & float   &   deg      &   Right ascension (J2000)                                 \\
DEdeg                    & float   &   deg      &   Declination (J2000)                                     \\
RUWE                   & float   &   -        &   Renormalized unit weight error                          \\
prob                   & float   &   -        &   Membership probability                                  \\
Bin\_phot.\_flag           &   int  &    -     &   Photomtric binary flag. Binary = 2, MS single = 1, Unkown = 0  \\
Bin\_ruwe\_flag            &   int  &    -     &   RUWE binary flag (RUWE $\geq$ 1.4), Yes = 1, No =0\\
SB\_flag                   &   int  &    -     & Spectroscopic binary\\
SB\_ref                   &   string  &    -     & Flag of the SB\_flag source\\
BSS\_flag                  &   int  &    -     & Blue straggler star candidate flag. Yes =1, No=0 \\
 
Prot                     & float  &   days   &  Rotation period \\
e\_Prot                     & float  &   days   &  uncertainty of rotation period \\
PSNR                     & float  &    -     &  Signal-to-noise (SNR) of the rotation period\\
Ref\_flag$^a$                     & int  &   -   &  Flag of the Prot source\\
\hline
\\
\end{tabular}
\tablecomments{$^b$Flag of the Prot source: [0]: This work; [1]: \citet{Meibom2009}; [2]: \citet{Hartman2010};  [3]: \citet{Hartman2011AJ....141..166H}; [4]: \citet{Delorme2011MNRAS.413.2218D}; [5]: \citet{Nardiello2015}; [6]: \citet{Libralato2016}; [7]: \citet{Curtis2019ApJ...879...49C}; [8]: \citet{Soares2020}; [9]: \citet{Jeffries2021}; [10]: \citet{Dungee2022ApJ...938..118D}. This table is available in its entirety in machine-readable from.}
\end{table*}

\subsection{Period-Color Diagram}
\begin{figure*}
\centering
\gridline{\fig{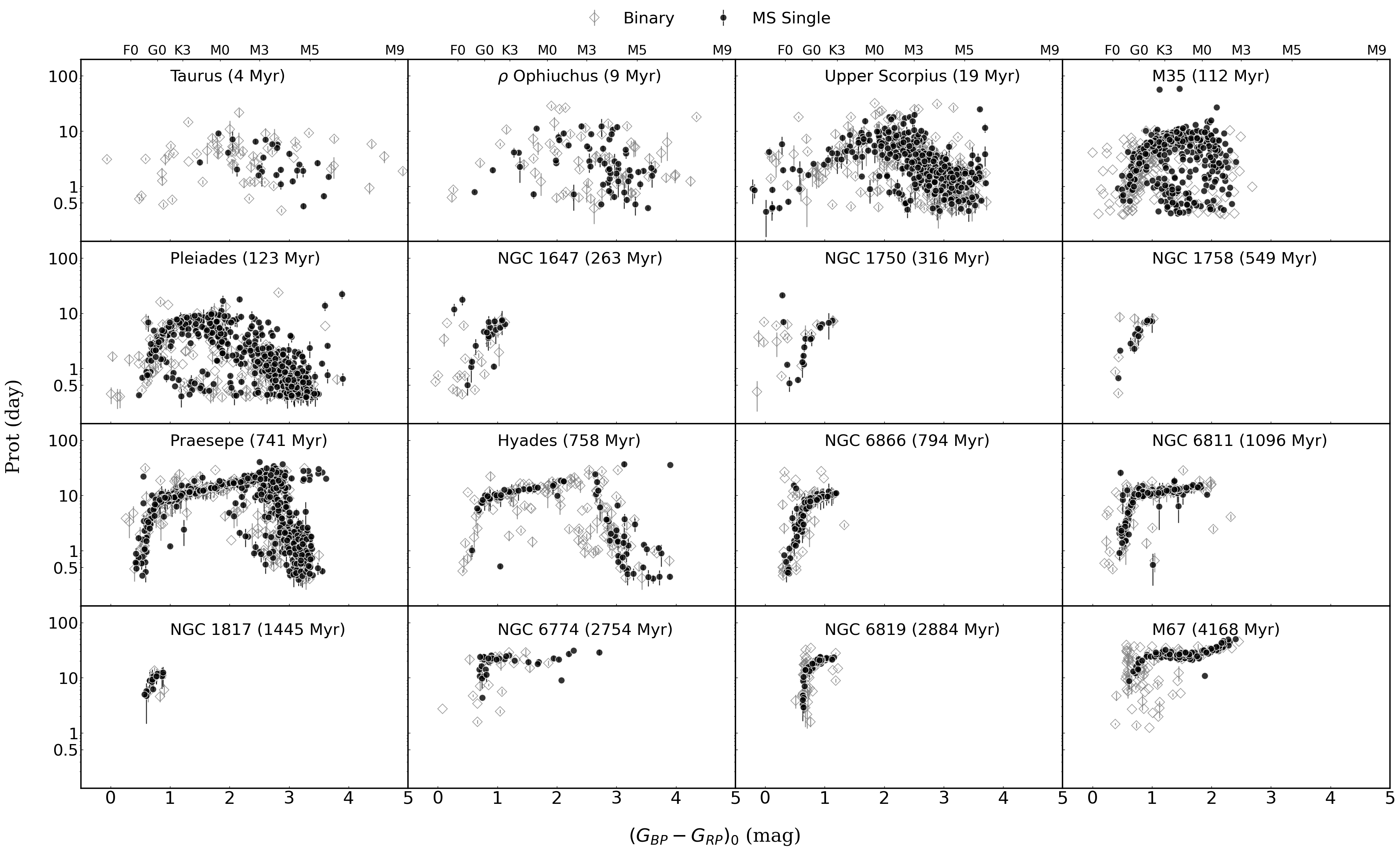}{1\textwidth}{}
          }
\caption{The color-period diagrams for 16 OCs. Black dots are MS single stars, gray diamonds are binary candidates.
\label{fig:Prot_distribution}}
\end{figure*}

\indent We construct a timeline benchmark $P_{\rm rot}$ data to characterize rotational behavior with ages. Fig.\ref{fig:Prot_distribution} shows the color-period diagrams (CPDs) of 16 OCs. It can be seen that the rotation distributions in different OCs at different ages reveals notable features:\\ 
\begin{itemize}
\item For the youngest clusters (i.e. Taurus, $\rho$ Ophiuchus), the sequences of binaries and single stars are insignificant, with the populations displaying large dispersion of $P_{\rm rot}$, ranging from 0.5 to 20 days. After $\sim$20 Myrs (i.e. Upper Scorpius), the binaries share similar sequences with single stars, while the features for binaries are more scattered in $P_{\rm rot}$. 
\item Rotation sequence tends to be clear at the age of Upper Scorpius. For FGK-type stars, $P_{\rm rot}$ increase monotonically with color (defined as slow rotator sequence), while the changing trend shows in opposite direction for M-type stars. 
\item For M35 and Pleiades, besides the similar sequence as that shown for Upper Scorpius, there is a separated sequence in K- and M-type stars occupied with the rapid rotators with their $P_{\rm rot}$ decreasing monotonically with color (defined as rapid rotator sequence). The slow and rapid rotator sequences have been initially displayed in \citet{Barnes2003}. As the age of the cluster increases, the period-color relation of stars in the slow rotator sequence becomes increasingly tighter, and stars in the rapid rotator sequence gradually transit to the slow sequence.

\item On the slow rotator sequence, the stars with different spectral types (i.e. FG-, K-, and M-type) show different spin-down rates, after the age of M35, which may indicate different physical mechanisms that drive their angular momentum evolution. Specifically, for the K-type stars of NGC 6811, the $P_{\rm rot}$ values overlap with Praesepe's, indicating that the spin-down rate of K-type stars is temporarily stalled. The phenomenon is also reported by \citet{Curtis2019ApJ...879...49C} and \citet{Douglas2019}. The K-type stars of NGC 6774 and M67 also exhibit similar rotational periods.
\end{itemize}

\subsection{An Empirical Period-Age-Color Relation}
\begin{figure*}
\centering
\includegraphics[height=7.5cm,width=16cm]{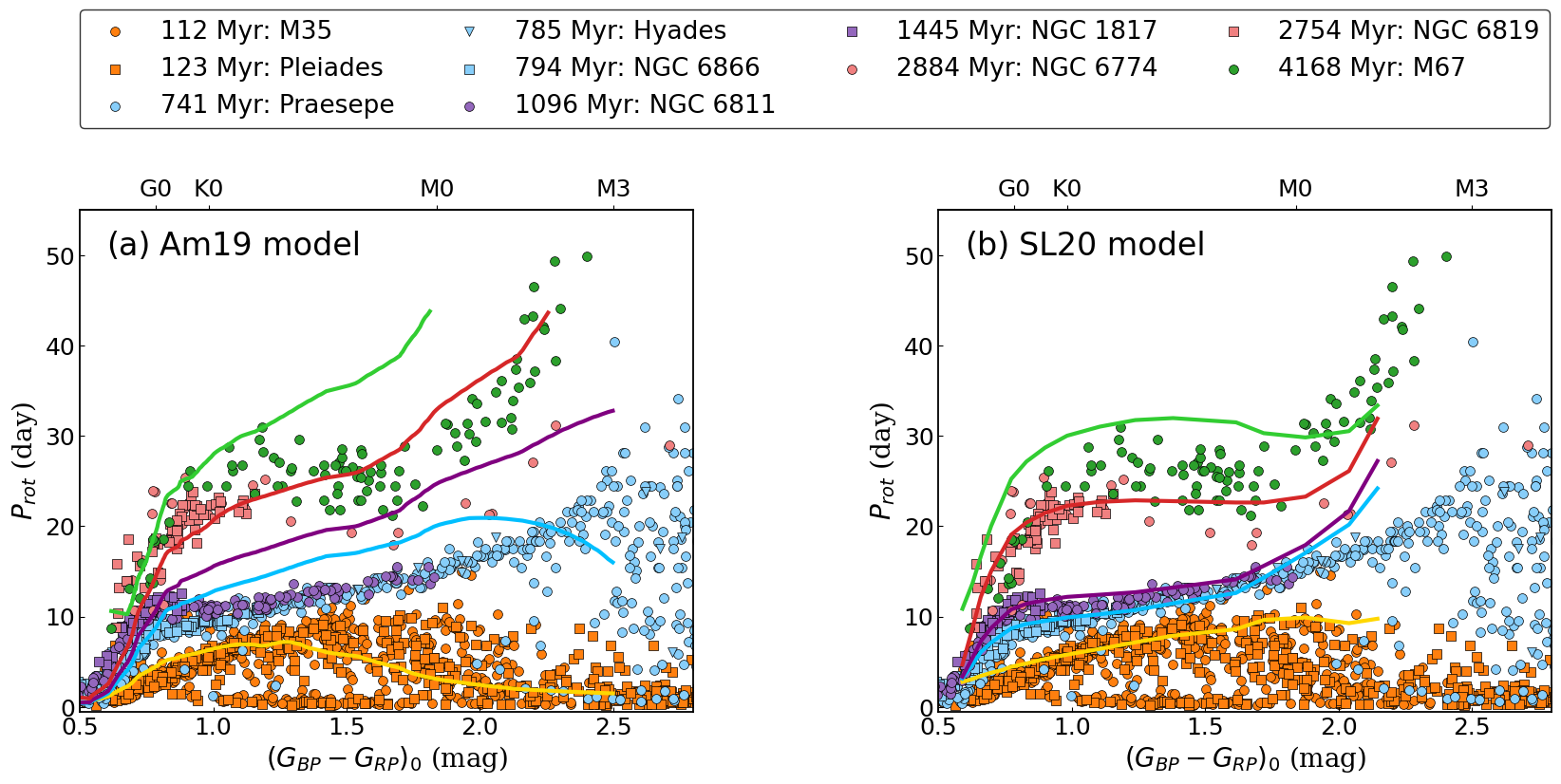}
\caption{Comparisons between the $P_{\rm rot}$ data for benchmark populations with \citet{Amard2019A&A...631A..77A} model (a) and \citet{Spada2020A&A...636A..76S} model (b). The ages corresponding to the solid lines on panel (a) and (b) are 120 Myr, 700 Myr, 1 Gyr, 2.5 Gyr, 4 Gyr from bottom to top. Clusters excluded from benchmark populations either have few $P_{\rm rot}$ data or are too young to show color-period relations.\label{fig:slow_seq}}
\end{figure*}

\begin{figure}[ht!]
\plotone{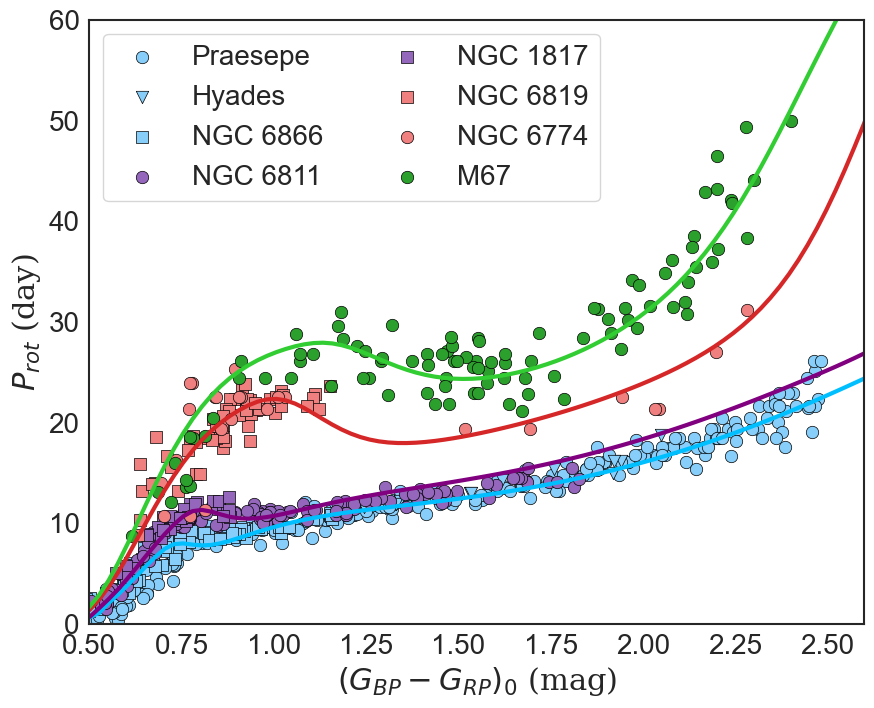}
\caption{$P_{\rm rot}$ vs. $(G_{BP}-G_{RP})_0$ of clusters with ages older than Praesepe. Solid lines are fits to the data. \label{fig:gyro_fit}}.
\end{figure}

\indent We use $P_{\rm rot}$ of single stars in a timeline to limit the gyrochronological models. The model shown in Fig.\ref{fig:slow_seq} (a) is the \citet{Amard2019A&A...631A..77A} model (hereafter Am19), which implements the wind braking law, persented by \citet{Matt2015ApJ...799L..23M}, into the
stellar models. It can be found that this model matches the observations in the entire color range for M35 and Pleiades. For clusters older than Praresepe, only the F- and G -type stars are compatible with the slow rotator sequence of model predictions, while the $P_{\rm rot}$ of the K- and M-type stars are all overestimated. The model shown in Fig.\ref{fig:slow_seq} (b) is the \citet{Spada2020A&A...636A..76S} model (hereafter SL20), with the key ingredients being the scalings of the wind braking law and the rotational coupling timescale with stellar mass. The predictions produced by this model are in good agreement with the observations of FG-type stars. However, for K-type stars, the predictions are only applicable to those in M35, Pleiades, Praespe and NGC 6811, while for NGC 6774 and M67 do not fit in well with the observations. Additionally, for M-type stars, the predictions by this model are longer than the observations for clusters older than Praesepe.\\
\indent As demonstrated, these models do not fit in well with the observational features in some cases, especially for the low-mass stars in relatively old clusters. Thus, we attempt to construct an empirical relation for low-mass single stars on the slow rotator sequence. Inspired by the empirical Praesepe-calibrated gyrochronology relation from \citet{Angus2019AJ....158..173A}, we consider the effect of stalled spin-down in the model. The period-age-color relation for stars with 0.5 $\leq$ ($G_{BP}$-$G_{RP}$)$_0$ $\leq$ 2.5 is defined as:
\begin{equation}
\begin{aligned}
\log _{10}\left( P_{rot} \right) \,\,=& c_A\log_{10} \left( t \right) +\sum_{n=0}^4{c_n\left[ \log _{10}\left( G_{BP}-G_{RP} \right)_0 \right]}\\
&+\log _{10}\left( dP_{rot} \right) 
\end{aligned}
\end{equation}
where $t$ is age in years, and $dP_{rot}$ is the rotation period in the stall time with
\begin{equation}
\begin{aligned}
dP_{rot}\ =\ \phi \left( t \right) *\frac{\left( t_2-t_1 \right)}{2}*\frac{\sqrt{2\pi}}{2}*e^{\left( \frac{1}{2|skw|}*\left( P_1-P_2 \right) *samp \right)}
\end{aligned}
\end{equation}
where $skw$ is the skewness of the stall vs. acceleration, $samp$ is the stall amplitude, $\phi \left( t \right)$ is the normalization of the skewnorm dist of $t$, $t_1$ and $t_2$ are starting and ending times to stall that estimated by 
\begin{equation}
t_k = 10^{(\log _{10}\left( P_{rot,k} \right) - \sum_{n=0}^4{c_n\left[ \log _{10}\left( G_{BP}-G_{RP} \right)_0 \right]})/c_A}
\end{equation}
\begin{equation}
P_{rot,k} = Ro_{k}\times \tau_c
\end{equation}

where $k$ = (1, 2), $Ro$ is Rossby number, $\tau_c$ is the convective turnover time scale and is calculated with Equation 36 in \citet{Cranmer2011ApJ...741...54C}. 
The best-fit coefficient values are listed in Table \ref{tab:table6}, and show that the star remain stalled spin-down with a $Ro$ ranging from 0.604 to 0.963. We display the comparison between the observational data with our empirical rotational isochrones in Fig.\ref{fig:gyro_fit}. It can be found that the rotational isochrones given by our relation fit in well with the observations. This relation applies to stars with age between 700-4000 Myr and spectral types FGK and early M. For convenience, we provide codes for generating rotational isochrones and deriving ages at \href{https://github.com/longliu-git/gyroage.git}{github.com/longliu-git/gyroage.git}.\\

\begin{table}[h!t]
\scriptsize
\centering
\caption{Coefficient values for fitting.}\label{tab:table6}
\begin{tabular}{lc}
\hline
\hline
\multicolumn{1}{c}{Coefficient} &
\multicolumn{1}{c}{Value}\\
  \hline
$c_0$ & -2.710 $\pm$ 0.010\\
$c_1$ & 0.526 $\pm$ 0.004\\
$c_2$ & -3.301 $\pm$ 0.015\\
$c_3$ & 17.536 $\pm$ 0.183\\
$c_4$ & -26.299 $\pm$ 0.888\\
$c_5$ & 21.354 $\pm$ 1.219\\
$c_A$ & 0.431 $\pm$ 0.001\\
$Ro_1$ & 0.604 $\pm$ 0.397\\
$Ro_2$ & 0.963 $\pm$ 0.820\\
$skw$ & -0.053 $\pm$ 0.194\\
$samp$ & 0.001 $\pm$ 0.141\\
\hline
\end{tabular}

\end{table}

\section{Conclusions}\label{sec:conclusion}
Based on Gaia DR3 and Kepler/K2 data, we present a catalog of 16 OCs in time-series with membership, binary, stellar rotation, and provide an updated view of rotation-age-color relation. This catalog includes 20,160 memberships, among which the 4,381 targets are flagged as binary candidates and the rotation periods of the 5,467 targets are determined. We further examine the rotation-age-color relations over time. Our main conclusions are as follows:\\
\indent 1. Combining the metallicity from the spectral data with the Gaia DR3, we deduce the age, distance modulus and reddening of 16 OCs by isochrone fitting. In addition, we estimate cluster astrometric and spatial structure properties from identified members.\\
\indent 2. The fractions of binaries are from 9\% to 44\% in 16 OCs. Moreover, there are 49 blue straggler candidates being detected in four old open clusters (NGC 1817, NGC 774, NGC 6819 and M67). \\
\indent 3. To characterize the distributions of rotation with age and color, we construct period-color diagrams of 16 time-series open clusters. The period-color diagrams reveal that: (i) the binary and single stars exhibit on similar rotator sequences, while binary stars are more scattered in rotation; (ii) the dependence of the $P_{\rm rot}$ on the color (or mass) has already been established at the age of Upper Scorpius; (iii) After the age of M35, stars with varying spectral types (i.e. FG-, K-, and M-types) exhibit diverse spin-down rates, suggesting that the physical mechanisms driving angular momentum loss may be different. Moreover, the spin-down in K-type stars are temporarily stalled for the relatively old cluster.\\
\indent 4. The Am19 and SL20 models for slow rotator sequence do not fit in well with stars of K- and M-types between Praesepe and M67. Therefore, we develop an empirical rotation-age-color relation based on the observational features, which suggests that the star remain stalled spin-down within $Ro$ = [0.604, 0.963]. This relation allows the estimation of ages from 700 to 4000 Myr with colors of 0.5 $<$ $(G_{BP}-G_{RP})_0$ $<$ 2.5.\\

\begin{acknowledgments}
 \indent This work is supported by the Joint Research Fund in Astronomy (U2031203) under cooperative agreement between the National Natural Science Foundation of China (NSFC) and Chinese Academy of Sciences (CAS), National Key R$\&$D Program of China No. 2019YFA0405503, and NSFC grants (12090040, 12090042). J.-H.Z. acknowledges support from NSFC grant No. 12103063 and from China Postdoctoral Science Foundation funded project (grant No. 2020M680672). This work is partially supported by the Scholar Program of Beijing Academy of Science and Technology (DZ:BS202002) and the CSST project.\\
 \indent This work has made use of data from the European Space Agency (ESA) mission Gaia (\url{https://www.cosmos.esa.int/gaia}), processed by the Gaia Data Processing and Analysis Consortium (DPAC, \url{https://www.cosmos.esa.int/web/gaia/dpac/consortium}). Funding for the DPAC has been provided by national institutions, in particular the institutions participating in the Gaia Multilateral Agreement.\\
 \indent Some of the data presented in this paper were obtained from the Mikulski Archive for Space Telescopes (MAST) at the Space Telescope Science Institute. The specific observations analyzed can be accessed via \dataset[DOI: 10.17909/T93W28]{https://doi.org/10.17909/T93W28} and \dataset[DOI: 10.17909/T9059R]{https://doi.org/10.17909/T9059R}.\\

\end{acknowledgments}
\facilities{Gaia \citep{Gaia2022arXiv220800211G}, Kepler/K2 \citep{Borucki2010Sci...327..977B,Howell2014PASP..126..398H}}
\software{lightkurve\citep{Lightkurve2018ascl.soft12013L}, HDBSCAN\citep{Campello,McInnes2017JOSS....2..205M}
          }

\appendix

\section{Appendix Plots of cluster members}\label{sec:AppendixA}
\setcounter{table}{0}
\renewcommand\thetable{A.\arabic{table}}
\setcounter{figure}{0}
\renewcommand\thefigure{A.\arabic{figure}}

\begin{figure}
\includegraphics[width=1\textwidth]{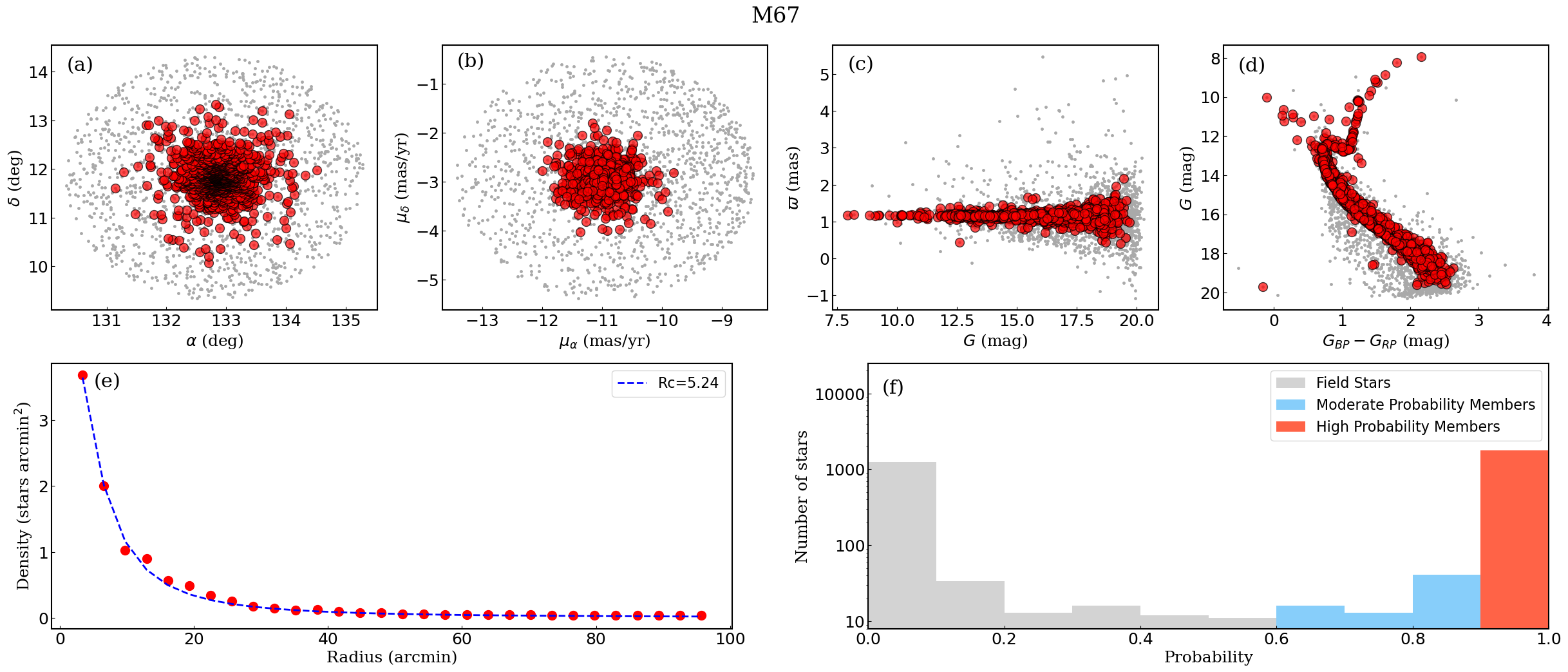}
\caption{The distributions of cluster members for M67. (a) Spatial distribution. (b) VPD. (c) Distribution of $G$ magnitude with parallax. (d) CMD. For (a)-(d), $Sample$ $sources$ are denoted by grey dots and cluster members are denoted by red points. (e) Calculation of core radius by ﬁtting radial density with King’s surface density proﬁle. (f) Distribution of the membership probabilities. (The complete figure sets (13 images) is available.)
 \label{fig:Memb_1}}
\end{figure}

\indent Plots showing properties of likely cluster members for each of the 13 OCs in age order, which contain spatial distributions, VPD, distributions of $G$ magnitude with parallax, CMDs, distributions of RDP, distributions of the membership probabilities. We take M67 as an example in Fig.\ref{fig:Memb_1}, the complete figure sets are available in the online journal.

\figsetstart
\figsetnum{12}
\figsettitle{The distributions of cluster members}

\figsetgrpstart
\figsetgrpnum{1.1}
\figsetgrptitle{M35}
\figsetplot{FigA1_M35.png}
\figsetgrpnote{The distributions of cluster members for M35}
\figsetgrpend

\figsetgrpstart
\figsetgrpnum{1.2}
\figsetgrptitle{Pleiades}
\figsetplot{FigA1_Pleiades.png}
\figsetgrpnote{The distributions of cluster members for Pleiades}
\figsetgrpend

\figsetgrpstart
\figsetgrpnum{1.3}
\figsetgrptitle{NGC 1647}
\figsetplot{FigA1_NGC1647.png}
\figsetgrpnote{The distributions of cluster members for NGC 1647}
\figsetgrpend

\figsetgrpstart
\figsetgrpnum{1.4}
\figsetgrptitle{NGC 1750}
\figsetplot{FigA1_NGC1750.png}
\figsetgrpnote{The distributions of cluster members for NGC 1750}
\figsetgrpend

\figsetgrpstart
\figsetgrpnum{1.5}
\figsetgrptitle{NGC 1758}
\figsetplot{FigA1_Pleiades.png}
\figsetgrpnote{The distributions of cluster members for NGC 1758}
\figsetgrpend

\figsetgrpstart
\figsetgrpnum{1.6}
\figsetgrptitle{Hyades}
\figsetplot{FigA1_Hyades.png}
\figsetgrpnote{The distributions of cluster members for Hyades}
\figsetgrpend

\figsetgrpstart
\figsetgrpnum{1.7}
\figsetgrptitle{Praesepe}
\figsetplot{FigA1_Praesepe.png}
\figsetgrpnote{The distributions of cluster members for Praesepe}
\figsetgrpend

\figsetgrpstart
\figsetgrpnum{1.8}
\figsetgrptitle{NGC 6866}
\figsetplot{FigA1_NGC6866.png}
\figsetgrpnote{The distributions of cluster members for NGC 6866}
\figsetgrpend

\figsetgrpstart
\figsetgrpnum{1.9}
\figsetgrptitle{NGC 6811}
\figsetplot{FigA1_NGC6811.png}
\figsetgrpnote{The distributions of cluster members for NGC 6811}
\figsetgrpend
\figsetgrpstart
\figsetgrpnum{1.10}
\figsetgrptitle{NGC 1817}
\figsetplot{FigA1_NGC1817.png}
\figsetgrpnote{The distributions of cluster members for NGC 1817}
\figsetgrpend

\figsetgrpstart
\figsetgrpnum{1.11}
\figsetgrptitle{NGC 6819}
\figsetplot{FigA1_NGC6819.png}
\figsetgrpnote{The distributions of cluster members for NGC 6819}
\figsetgrpend

\figsetgrpstart
\figsetgrpnum{1.12}
\figsetgrptitle{NGC 6774}
\figsetplot{FigA1_NGC6774.png}
\figsetgrpnote{The distributions of cluster members for NGC 6774}
\figsetgrpend


\figsetend

\section{Appendix Binary members and BSS in CMDs}\label{sec:AppendixB}
\setcounter{figure}{0}
\renewcommand\thefigure{B.\arabic{figure}}
Fig \ref{fig:binary} shows CMDs of high probability members ($P > 0.9$) and the stars was flagged as candidate binaries and BSS. 
\begin{figure*}
\addtocounter{subfigure}{6} %
\centering
\includegraphics[width=1\textwidth]{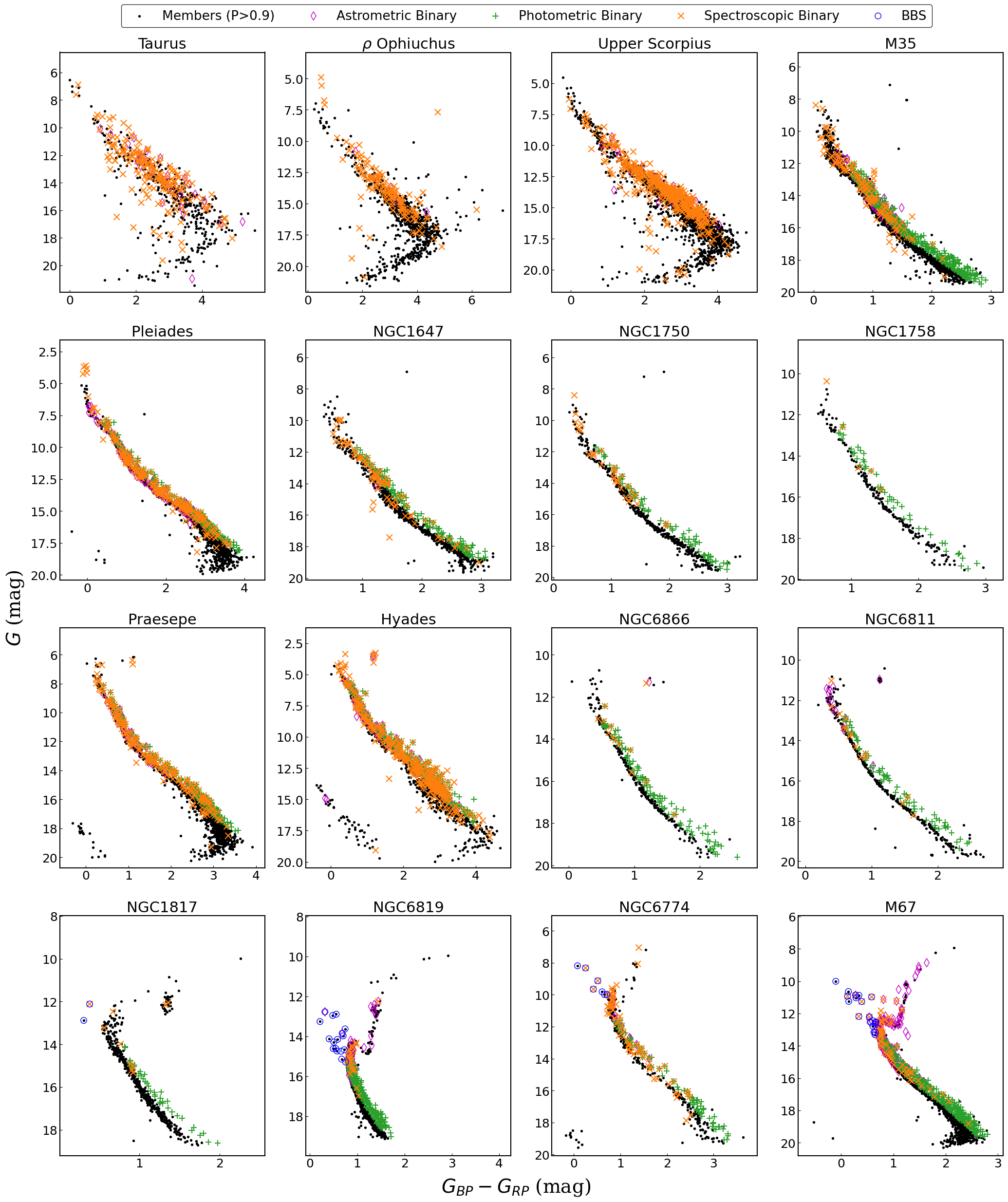}
\caption{CMDs of 16 OCs. The black dots are single stars. The magenta diamonds are stars flagged as candidate astrometric binaries with RUWE $\leq$ 1.4. The green pluses are stars flagged as candidate photometric binaries. The orange crosses are stars flagged as candidate spectroscopic binaries. The blue circles are stars flagged as candidate BBS.
\label{fig:binary}}
\end{figure*}

\bibliography{ref.bib}{}
\bibliographystyle{aasjournal}

\end{document}